\documentclass[hidelinks,11pt]{article}
\usepackage{natbib,amsmath,amsthm,amssymb,color,graphicx}
\usepackage{times}
\usepackage{epstopdf}
\usepackage{multirow}
\usepackage{hyperref}

\usepackage[usenames,dvipsnames]{xcolor}

\newcommand{\betaIBP}{\gamma}
\newcommand{\alphaIBP}{\alpha}
\newcommand{\bkappa}{b^\kappa}
\newcommand{\ckappa}{c^\kappa}
\newcommand{\bsigma}{b^\sigma}
\newcommand{\csigma}{c^\sigma}
\newcommand{\aalpha}{a^\alpha}
\newcommand{\balpha}{b^\alpha}
\newcommand{\abeta}{a^\betaIBP}
\newcommand{\bbeta}{b^\betaIBP}
\newcommand{\ThetaD}[1]{\Theta_{#1}}
\newcommand{\facloadrref}{{\mathbf B}}

\newcommand{\rtrue}{r_{\mbox{\tiny \rm true}}}

\newcommand{\tumS}{S}
\newcommand{\GLTAR}{\mbox{\bf GLT-AR}}

\newcommand{\AR}{\mbox{\bf AR}}

\newcommand{\Rs}{s}
\newcommand{\RD}[2]{\mbox{RD} (#1,#2)}
\newcommand{\CR}[2]{\mbox{CR} (#1,#2)}

\newcommand{\rankl}{z}

\newcommand{\CountAR}{3579}

\newcommand{\Gammainv}[1]{\mathcal{G}^{-1} \left(#1\right)}
\newcommand{\Gammad}[1]{ \mathcal{G}\left(#1\right)}

\newcommand{\Normal}[1]{ \mathcal{N}\left(#1\right)}
\newcommand{\Normult}[2]{ \mathcal{N} _{#1}\left(#2\right)}

\newcommand{\Bincoeftext}[2]{{\small \left(\begin{array}{c}   #1 \\[-3mm] #2  \end{array}\right)}}
\newcommand{\Bincoefsmall}[2]{ #1 ! /(#2 ! (#1 -#2 ) !)  }

\newcommand{\Betadis}[1]{\mathcal{B}\left(#1\right)}

\newcommand{\Real}{\mathbb{R}}

\newcommand{\Sm}{{\mathbf S}}
\newcommand{\Rm}{\mathbf{R}}

\newcommand{\ym}{{\mathbf y}}             %

\newcommand{\om}{\Omega}                  %
\newcommand{\Vary}{{\mathbf \om}}         %
\newcommand{\dimy}{m}                     %
\newcommand{\load}{\beta}                 %
\newcommand{\facload}{\boldsymbol{\load}} %
\newcommand{\betatrue}{\boldsymbol{\load}_0} %
\newcommand{\nfac}{k}                     %
\newcommand{\nspu}{s}
\newcommand{\fac}{f}                      %
\newcommand{\facm}{{\mathbf \fac}}        %
\newcommand{\error}{\epsilon}             %
\newcommand{\errorm}{\boldsymbol{\error}} %
\newcommand{\Vare}{{\mathbf \Sigma}}      %

\newcommand{\Varetrue}{\Vare_0}      %
\newcommand{\Cov}[1]{\mbox{\rm Cov}(#1)}
\newcommand{\idiov}{\sigma^2}             %
\newcommand{\Pm}{{\mathbf P}}             %
\newcommand{\rotG}{{\mathbf G}}

\newcommand{\Qm}{{\mathbf Q}}             %
\newcommand{\Mm}{{\mathbf M}}             %
\newcommand{\Dm}{{\mathbf D}}             %
\newcommand{\Lm}{{\mathbf L}}             %
\newcommand{\Bm}{{\mathbf B}}             %
\newcommand{\Cm}{{\mathbf C}}             %
\newcommand{\Gm}{{\mathbf G}}             %
\newcommand{\Tm}{{\mathbf T}}             %
\newcommand{\deltav}{\boldsymbol{\delta}} %

\newcommand{\bm}{{\mathbf b}}             %
\newcommand{\Pim}{\boldsymbol{\Pi}}       %
\newcommand{\bP}{\Pm}

\newcommand{\ones}{{\mathbf{1}}}

\newcommand{\facloadtilde}{\tilde{\facloadtrue}} %
\newcommand{\Varetilde}{\tilde{\Vare}_0}      %

\renewcommand{\facloadtilde}{\facload} %
\renewcommand{\Varetilde}{\Vare}
\newcommand{\Am}{{\mathbf A}}         %
\newcommand{\Diag}[1]{\mbox{\rm Diag}\!\left(#1\right)} %
\newcommand{\dimmat}[2]{#1\times #2}  %
\newcommand{\bfz}{{\mathbf{0}}}         %
\newcommand{\bfzmat}{{\mathbf{O}}}      %
\newcommand{\identm}{{\mathbf I}}       %
\newcommand{\identy}[1]{{\identm}_{#1}} %
\newcommand{\Probsym}{\mbox{\rm Pr}}    %
\newcommand{\Prob}[1]{\Probsym (#1)}    %
\newcommand{\Ew}[1]{\mbox{\rm E}(#1)}   %
\newcommand{\rank}[1]{\mbox{\rm rk}\,(#1)}  %
\newcommand{\trans}[1]{#1^{\top}}         %
\newcommand{\indic}[1]{\mathbb{I} (#1)}

\newcommand{\loadtrue}{\Lambda}                 %
\newcommand{\loadtruetilde}{\tilde{\Lambda}}                 %
\renewcommand{\loadtruetilde}{\beta}
\newcommand{\facloadtrue}{\boldsymbol{\loadtrue}} %
\newcommand{\nfactrue}{r}
{\begin{figure}[t!]
		\begin{center}
			\scalebox{#4}{\includegraphics{#3.eps}}
			\caption{#1}\label{#2}}
		{\end{center}\end{figure}}

{\begin{figure}[t!]
		\begin{center}
			\scalebox{#5}{\includegraphics{#3.eps}}
			\hspace*{1cm}
			\scalebox{#5}{\includegraphics{#4.eps}}
			\caption{#1}\label{#2}
		}{\end{center}\end{figure}}

{\begin{figure}[t!]
		\begin{center}
			\begin{tabular}{ccc}
				\scalebox{#6}{\includegraphics{#3.eps}}
				& %
				\scalebox{#6}{\includegraphics{#4.eps}}
				&
				\scalebox{#6}{\includegraphics{#5.eps}}
			\end{tabular}
			\caption{#1}\label{#2}
		}{\end{center}\end{figure}}

\theoremstyle{plain}
\newtheorem{thm}{Theorem}%
\newtheorem{cor}[thm]{Corollary}
\newtheorem{lem}[thm]{Lemma}

\theoremstyle{definition}

\newtheorem{defn}{Definition}

\usepackage{tikz}
\usetikzlibrary{positioning, shapes.geometric}

\def\stepsize{3mm}
\colorlet{covercolor}{green!80!black}

\tikzset{nonzero/.style={draw, circle, color=white, fill=blue!90!black, inner sep=0.5mm}}
\tikzset{leading/.style={draw, isosceles triangle, isosceles triangle apex angle=60, color=white, fill=red!90!black, inner sep=0.5mm, rotate=-30}}
\tikzset{axis label/.style={}}
\tikzset{matrix box/.style={ultra thin, color=gray}}

\begin{document}
	
\title{When it counts - Econometric identification of the basic factor model based on GLT structures}

\author{Sylvia Fr\"uhwirth-Schnatter\footnote{Department of Finance, Accounting, and Statistics, WU Vienna University of Economics and Business, Austria. Email: {\tt sfruehwi@wu.ac.at}} \and
 Darjus Hosszejni\footnote{Department of Finance, Accounting, and Statistics, WU Vienna University of Economics and Business, Austria. Email: {\tt darjus.hosszejni@wu.ac.at}}
\and Hedibert Freitas Lopes\footnote{School of Mathematical and Statistical Sciences, Arizona State University, Tempe, USA \& Insper Institute of Education and Research, S\~ao Paulo, Brazil. Email: {\tt hedibertfl@insper.edu.br}}, \,
}

\maketitle

\begin{abstract}
Despite the  popularity of factor models with sparse loading matrices, little attention has been given to formally address  identifiability of these models beyond standard rotation-based identification such as the positive lower triangular (PLT) constraint.
To fill this gap, we review the advantages of variance identification in sparse factor analysis and introduce the generalized lower triangular (GLT) structures.
We show that the GLT assumption is an improvement over PLT without compromise: GLT is also unique but, unlike PLT, a non-restrictive assumption.
Furthermore, we provide a simple counting rule for variance identification under GLT structures, and we demonstrate that within this model class the unknown number of common factors can be recovered in an exploratory factor analysis.
Our methodology is illustrated for simulated data in the context of post-processing posterior draws in Bayesian sparse factor analysis.

\end{abstract}

\vspace{0.5cm}
\vspace{0.5cm}

\centerline{JEL classification: C11, C38, C63}

\section{Introduction}

Ever since the pioneering work of \citet{thu:vec,thu:mul}, factor analysis has been a popular method  to model the covariance matrix $\Vary$ of correlated, multivariate observations $\ym_t$ of dimension $\dimy$, see  e.g.\ \citet{and:int} for a comprehensive review.
Assuming $\nfactrue$  uncorrelated factors, the  basic factor model yields the representation  $\Vary=  \facloadtrue  \trans{\facloadtrue } + \Varetrue $, with a   $\dimmat{\dimy}{\nfactrue}$ factor loading matrix   $\facloadtrue$  and  a diagonal matrix $\Varetrue $.
The considerable reduction of the number of   parameters  compared to the $\dimy(\dimy+1)/2$ elements of an unconstrained  covariance matrix $\Vary$ is the main  motivation for applying  factor models to covariance estimation,
especially if $\dimy$ is large; see, among many others, \citet{fan-etal:hig_je} in finance and \citet{for-etal:ope} in economics. In addition, shrinkage estimation has been shown to lead to very efficient covariance estimation,
see, for example, \citet{kas:spa} in Bayesian factor analysis and  \citet{led-wol:pow} in a non-Bayesian context.

In numerous applications,  factor analysis reaches beyond covariance modelling. From the very beginning, the goal
of factor analysis has been to extract the underlying loading matrix $\facloadtrue$ to understand the driving forces behind the observed correlation between the features, see e.g.\ \citet{owe-wan:bic} for a recent review.
However,  also in this setting, the only source of information is the observed covariance of the data,
making the decomposition of the covariance matrix  $\Vary$ into the cross-covariance matrix $\facloadtrue  \trans{\facloadtrue}$ and the variance $ \Varetrue $  of the idiosyncratic errors more challenging than estimating only $\Vary$ itself.

A huge literature, dating  back to \citet{koo-rei:ide} and \citet{rei:ide}, has addressed this problem of identification
which can be resolved only by imposing additional structure on the factor model. \citet{and-rub:sta} considered  identification as a two-step procedure, namely identification of $\Varetrue$ from $\Vary$ (variance identification) and subsequent identification of $\facloadtrue$ from  $\facloadtrue  \trans{\facloadtrue }$ (solving rotational invariance). The most popular  constraint  in econometrics, statistics and machine learning for solving rotational invariance is to consider positive lower triangular loading matrices, see e.g.\ \citet{gew-zho:mea,wes:bay_fac,lop-wes:bay}, albeit other strategies have been put forward, see e.g.\ \citet{neu:ide}, \citet{bai-ng:pri}, \citet{ass-etal:bay}, \citet{cha-etal:inv}, and \citet{wil:ide}. Only a few papers have addressed variance identification \citep[e.g.][]{bek:ide} and to the best of our knowledge so far no structure has been put forward that simultaneously addresses both  identification problems.

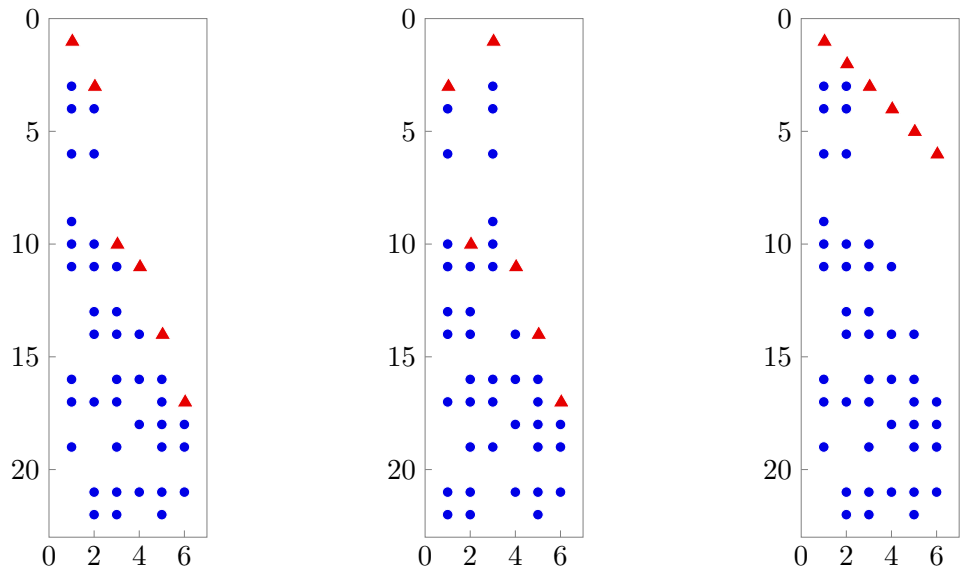
\begin{figure}[t]
\centering
\begin{tikzpicture} [x=\stepsize, y=\stepsize, yscale=-1]
	\node[leading] at (1,1) {};
	\node[nonzero] at (1,3) {};
	\node[nonzero] at (1,4) {};
	\node[nonzero] at (1,6) {};
	\node[nonzero] at (1,9) {};
	\node[nonzero] at (1,10) {};
	\node[nonzero] at (1,11) {};
	\node[nonzero] at (1,16) {};
	\node[nonzero] at (1,17) {};
	\node[nonzero] at (1,19) {};
	\node[leading] at (2,3) {};
	\node[nonzero] at (2,4) {};
	\node[nonzero] at (2,6) {};
	\node[nonzero] at (2,10) {};
	\node[nonzero] at (2,11) {};
	\node[nonzero] at (2,13) {};
	\node[nonzero] at (2,14) {};
	\node[nonzero] at (2,17) {};
	\node[nonzero] at (2,21) {};
	\node[nonzero] at (2,22) {};
	\node[leading] at (3,10) {};
	\node[nonzero] at (3,11) {};
	\node[nonzero] at (3,13) {};
	\node[nonzero] at (3,14) {};
	\node[nonzero] at (3,16) {};
	\node[nonzero] at (3,17) {};
	\node[nonzero] at (3,19) {};
	\node[nonzero] at (3,21) {};
	\node[nonzero] at (3,22) {};
	\node[leading] at (4,11) {};
	\node[nonzero] at (4,14) {};
	\node[nonzero] at (4,16) {};
	\node[nonzero] at (4,18) {};
	\node[nonzero] at (4,21) {};
	\node[leading] at (5,14) {};
	\node[nonzero] at (5,16) {};
	\node[nonzero] at (5,17) {};
	\node[nonzero] at (5,18) {};
	\node[nonzero] at (5,19) {};
	\node[nonzero] at (5,21) {};
	\node[nonzero] at (5,22) {};
	\node[leading] at (6,17) {};
	\node[nonzero] at (6,18) {};
	\node[nonzero] at (6,19) {};
	\node[nonzero] at (6,21) {};

	\draw[matrix box] (0,0) rectangle (7,23);

	\begin{scope}[every node/.style={axis label, label distance=-1.5mm}]
		\node (v1) [label=left:{$0$}] at (0,0) {};
		\node (v2) [label=left:{$5$}] at (0,5) {};
		\node (v3) [label=left:{$10$}] at (0,10) {};
		\node (v4) [label=left:{$15$}] at (0,15) {};
		\node (v5) [label=left:{$20$}] at (0,20) {};
		\node (h1) [label=below:{$0$}] at (0,23) {};
		\node (h2) [label=below:{$2$}] at (2,23) {};
		\node (h3) [label=below:{$4$}] at (4,23) {};
		\node (h4) [label=below:{$6$}] at (6,23) {};
	\end{scope}

	\begin{scope}[matrix box]
		\draw (v2.center) -- +(0.3,0);
		\draw (v3.center) -- +(0.3,0);
		\draw (v4.center) -- +(0.3,0);
		\draw (v5.center) -- +(0.3,0);
		\draw (h2.center) -- +(0,-0.3);
		\draw (h3.center) -- +(0,-0.3);
		\draw (h4.center) -- +(0,-0.3);
	\end{scope}

	\begin{scope}[xshift=5cm]
		\node[leading] at (1,3) {};
		\node[nonzero] at (1,4) {};
		\node[nonzero] at (1,6) {};
		\node[nonzero] at (1,10) {};
		\node[nonzero] at (1,11) {};
		\node[nonzero] at (1,13) {};
		\node[nonzero] at (1,14) {};
		\node[nonzero] at (1,17) {};
		\node[nonzero] at (1,21) {};
		\node[nonzero] at (1,22) {};
		\node[leading] at (2,10) {};
		\node[nonzero] at (2,11) {};
		\node[nonzero] at (2,13) {};
		\node[nonzero] at (2,14) {};
		\node[nonzero] at (2,16) {};
		\node[nonzero] at (2,17) {};
		\node[nonzero] at (2,19) {};
		\node[nonzero] at (2,21) {};
		\node[nonzero] at (2,22) {};
		\node[leading] at (3,1) {};
		\node[nonzero] at (3,3) {};
		\node[nonzero] at (3,4) {};
		\node[nonzero] at (3,6) {};
		\node[nonzero] at (3,9) {};
		\node[nonzero] at (3,10) {};
		\node[nonzero] at (3,11) {};
		\node[nonzero] at (3,16) {};
		\node[nonzero] at (3,17) {};
		\node[nonzero] at (3,19) {};
		\node[leading] at (4,11) {};
		\node[nonzero] at (4,14) {};
		\node[nonzero] at (4,16) {};
		\node[nonzero] at (4,18) {};
		\node[nonzero] at (4,21) {};
		\node[leading] at (5,14) {};
		\node[nonzero] at (5,16) {};
		\node[nonzero] at (5,17) {};
		\node[nonzero] at (5,18) {};
		\node[nonzero] at (5,19) {};
		\node[nonzero] at (5,21) {};
		\node[nonzero] at (5,22) {};
		\node[leading] at (6,17) {};
		\node[nonzero] at (6,18) {};
		\node[nonzero] at (6,19) {};
		\node[nonzero] at (6,21) {};

		\draw[matrix box] (0,0) rectangle (7,23);

		\begin{scope}[every node/.style={axis label, label distance=-1.5mm}]
			\node (v1) [label=left:{$0$}] at (0,0) {};
			\node (v2) [label=left:{$5$}] at (0,5) {};
			\node (v3) [label=left:{$10$}] at (0,10) {};
			\node (v4) [label=left:{$15$}] at (0,15) {};
			\node (v5) [label=left:{$20$}] at (0,20) {};
			\node (h1) [label=below:{$0$}] at (0,23) {};
			\node (h2) [label=below:{$2$}] at (2,23) {};
			\node (h3) [label=below:{$4$}] at (4,23) {};
			\node (h4) [label=below:{$6$}] at (6,23) {};
		\end{scope}

		\begin{scope}[matrix box]
			\draw (v2.center) -- +(0.3,0);
			\draw (v3.center) -- +(0.3,0);
			\draw (v4.center) -- +(0.3,0);
			\draw (v5.center) -- +(0.3,0);
			\draw (h2.center) -- +(0,-0.3);
			\draw (h3.center) -- +(0,-0.3);
			\draw (h4.center) -- +(0,-0.3);
		\end{scope}
	\end{scope}

	\begin{scope}[xshift=10cm]
		\node[leading] at (1,1) {};
		\node[nonzero] at (1,3) {};
		\node[nonzero] at (1,4) {};
		\node[nonzero] at (1,6) {};
		\node[nonzero] at (1,9) {};
		\node[nonzero] at (1,10) {};
		\node[nonzero] at (1,11) {};
		\node[nonzero] at (1,16) {};
		\node[nonzero] at (1,17) {};
		\node[nonzero] at (1,19) {};
		\node[leading] at (2,2) {};
		\node[nonzero] at (2,3) {};
		\node[nonzero] at (2,4) {};
		\node[nonzero] at (2,6) {};
		\node[nonzero] at (2,10) {};
		\node[nonzero] at (2,11) {};
		\node[nonzero] at (2,13) {};
		\node[nonzero] at (2,14) {};
		\node[nonzero] at (2,17) {};
		\node[nonzero] at (2,21) {};
		\node[nonzero] at (2,22) {};
		\node[leading] at (3,3) {};
		\node[nonzero] at (3,10) {};
		\node[nonzero] at (3,11) {};
		\node[nonzero] at (3,13) {};
		\node[nonzero] at (3,14) {};
		\node[nonzero] at (3,16) {};
		\node[nonzero] at (3,17) {};
		\node[nonzero] at (3,19) {};
		\node[nonzero] at (3,21) {};
		\node[nonzero] at (3,22) {};
		\node[leading] at (4,4) {};
		\node[nonzero] at (4,11) {};
		\node[nonzero] at (4,14) {};
		\node[nonzero] at (4,16) {};
		\node[nonzero] at (4,18) {};
		\node[nonzero] at (4,21) {};
		\node[leading] at (5,5) {};
		\node[nonzero] at (5,14) {};
		\node[nonzero] at (5,16) {};
		\node[nonzero] at (5,17) {};
		\node[nonzero] at (5,18) {};
		\node[nonzero] at (5,19) {};
		\node[nonzero] at (5,21) {};
		\node[nonzero] at (5,22) {};
		\node[leading] at (6,6) {};
		\node[nonzero] at (6,17) {};
		\node[nonzero] at (6,18) {};
		\node[nonzero] at (6,19) {};
		\node[nonzero] at (6,21) {};

		\draw[matrix box] (0,0) rectangle (7,23);

		\begin{scope}[every node/.style={axis label, label distance=-1.5mm}]
			\node (v1) [label=left:{$0$}] at (0,0) {};
			\node (v2) [label=left:{$5$}] at (0,5) {};
			\node (v3) [label=left:{$10$}] at (0,10) {};
			\node (v4) [label=left:{$15$}] at (0,15) {};
			\node (v5) [label=left:{$20$}] at (0,20) {};
			\node (h1) [label=below:{$0$}] at (0,23) {};
			\node (h2) [label=below:{$2$}] at (2,23) {};
			\node (h3) [label=below:{$4$}] at (4,23) {};
			\node (h4) [label=below:{$6$}] at (6,23) {};
		\end{scope}

		\begin{scope}[matrix box]
			\draw (v2.center) -- +(0.3,0);
			\draw (v3.center) -- +(0.3,0);
			\draw (v4.center) -- +(0.3,0);
			\draw (v5.center) -- +(0.3,0);
			\draw (h2.center) -- +(0,-0.3);
			\draw (h3.center) -- +(0,-0.3);
			\draw (h4.center) -- +(0,-0.3);
		\end{scope}
	\end{scope}
\end{tikzpicture}
\caption{Left: ordered sparse GLT matrix with six factors.
	Center: one of the $2^6 \cdot 6$! corresponding unordered sparse GLT matrices.
	Right: a corresponding sparse PLT matrix, i.e.\ enforced non-zeros on the main diagonal.
	The pivot rows $(l_1, \ldots, l_6) =(1,3,10,11,14,17)$ are marked by triangles.
	Non-zero loadings are marked by circles, zero loadings are left blank.}
\label{figglt}
\end{figure}

In this work, we discuss a new  identification strategy  based on generalized lower triangular (GLT) structures, see Figure~\ref{figglt} for illustration. This concept was originally introduced  as part of an MCMC sampler for sparse Bayesian factor analysis where the number of factors is unknown  in the (unpublished) work of \citet{fru-lop:spa}. In the present paper, GLT structures are given a full and comprehensive mathematical treatment and are applied  in \citet{fru-etal:spa} to develop an efficient reversible jump MCMC (RJMCMC) sampler for sparse Bayesian factor analysis under very general shrinkage priors.
It  will be proven that GLT structures simultaneously address rotational invariance and variance identification  in factor models.  Variance identification relies on a counting rule for the number of non-zero elements in the loading matrix $\facloadtrue$, which is a sufficient condition that extends previous work by \citet{sat:stu}.

In addition, we will show that GLT structures are  useful in exploratory factor analysis where the factor dimension
$\nfactrue$ is unknown. Identification of the number of factors in applied factor analysis is a notoriously difficult problem,
with considerable ambiguity which method works best, be it BIC-type criteria
\citep{bai-ng:det2002}, marginal likelihoods \citep{lop-wes:bay}, techniques from Bayesian non-parametrics involving
infinite-dimensional factor models \citep{bha-dun:spa,roc-geo:fas,leg-etal:bay} or more heuristic procedures \citep{kau-sch:bay}.
Imposing an unordered GLT structure in exploratory factor analysis allows to identify the true loading matrix
$\facloadtrue$ and the matrix $\Varetrue $ and to easily spot all spurious columns in a possibly overfitting model.
This strategy underlies the RJMCMC sampler of \citet{fru-etal:spa} to estimate the number of factors.

The paper is structured as follows. Section~\ref{secmotivate} reviews the role of identification in factor analysis
using illustrative examples. Section~\ref{uniqload} introduces  GLT structures, proves identification for
sparse GLT structures and shows that any unconstrained loading matrix has a unique representation as a  GLT matrix.
Section~\ref{varidesp} addresses variance identification under GLT structures. %
Section~\ref{secEFA} discusses exploratory  factor analysis under
unordered GLT structures, while Section~\ref{secapll} presents an illustrative application.  Section~\ref{secconcluse} concludes.

\section{The role of identification in factor analysis} \label{secmotivate}

Let $\ym_t=\trans{(y_{1t}, \ldots,  y_{\dimy t})}$ be an observation vector of $\dimy$  measurements, which  is assumed to arise  from a multivariate normal distribution, $\ym_t  \sim \Normult{\dimy}{\bfz,\Vary}$, with zero mean and covariance matrix  $\Vary$.
In factor analysis, the correlation among the observations is assumed to be driven by a latent  $\nfactrue$-variate random variable $\facm_t=\trans{(\fac_{1t}, \ldots, \fac_{\nfactrue t})}$, the so-called common factors, through the following observation equation:
\begin{eqnarray}  \label{fac1}
\ym_t =  \facloadtrue \facm_t + \errorm_t,
\end{eqnarray}
where the $\dimmat{\dimy}{\nfactrue}$ matrix $\facloadtrue$ containing the
factor loadings $\loadtrue_{ij}$ is of full column rank, $\rank{\facloadtrue}=\nfactrue$,
equal to the factor dimension $\nfactrue$.
In the present paper, we
focus on the so-called basic factor model where
the vector $\errorm_t =\trans{(\error_{1t}, \ldots,  \error_{\dimy t})}$
accounts for independent, idiosyncratic variation of each measurement
and is distributed as  $\errorm_t
\sim \Normult{\dimy}{\bfz,\Varetrue} $,  with
$\Varetrue=\Diag{\idiov_{1},\ldots,\idiov_{\dimy}}$
being a  positive definite diagonal matrix.
The common factors are orthogonal, meaning that
$\facm_t  \sim  \Normult{\nfactrue}{\bfz,\identy{\nfactrue}},$ %
and  independent of %
$\errorm_t$. %
In this case, the observation equation (\ref{fac1})
implies the following covariance matrix $\Vary$,
when we integrate w.r.t.\ the latent common factors $\facm_t$:
\begin{eqnarray}
\Vary=     \facloadtrue  \trans{\facloadtrue } + \Varetrue .  \label{fac4}
\end{eqnarray}
Hence, all dependence among the measurements in  $\ym_t$
is explained through the latent common factors and
the off-diagonal elements
of $\facloadtrue  \trans{\facloadtrue }$ define
the marginal covariance between any two measurements $y_{i_1,t}$ and $y_{i_2, t}$:
\begin{eqnarray}
\Cov{y_{i_1,t}, y_{i_2, t}} = \facloadtrue _{i_1,\bullet} \trans{\facloadtrue _{i_2,\bullet}},
\label{fac5}
\end{eqnarray}
where $\facloadtrue _{i,\bullet}$ is the $i$th row of $\facloadtrue$. Consequently,
we will refer to  $\facloadtrue  \trans{\facloadtrue }$ as the cross-covariance matrix.
Since the number of factors, $\nfactrue$, is often considerably
smaller than the number  of measurements, $\dimy$, (\ref{fac4}) can be seen as a
parsimonious representation of the dependence between the measurements, often
with considerably fewer parameters in $\facloadtrue$ than the $\dimy(\dimy -1)/2$
off-diagonal elements in an unconstrained covariance matrix $\Vary$.

Since the factors $\facm_t$ are unobserved, the only information
available to estimate $\facloadtrue$ and $\Varetrue$
is the covariance matrix $\Vary$.  A rigorous approach toward identification of factor
models was first offered by \citet{rei:ide} and \citet{and-rub:sta}.
Identification  in the context of a basic factor model means the following. For any pair
$(\facloadtilde ,\Varetilde)$, where  $\facloadtilde$ is an $\dimy \times \nfactrue$ matrix and
$\Varetilde$ is a positive definite diagonal matrix,
that  satisfies (\ref{fac4}), i.e.:
\begin{eqnarray}
\Vary = \facloadtilde \trans{\facloadtilde} + \Varetilde
=     \facloadtrue  \trans{\facloadtrue } + \Varetrue,   \label{facide1}
\end{eqnarray}
it follows that  $\facloadtilde=\facloadtrue$ and
$\Varetilde =\Varetrue$.
Note that both parameter pairs imply the same Gaussian distribution $\ym_t \sim \Normult{\dimy}{\bfz,\Vary}$
for every possible realisation $\ym_t$.

\citet{and-rub:sta} considered  identification as a two-step procedure. The first step is
identification of the variance decomposition, i.e.\ identification of $\Varetrue$  from (\ref{fac4}),
which implies identification of   $\facloadtrue  \trans{\facloadtrue }$.
The second step is subsequent identification of $\facloadtrue$ from
$\facloadtrue  \trans{\facloadtrue }$, also know as solving the rotational invariance problem.
The  literature on factor analysis often
reduces identification of factor models to the second problem,
however as we will argue in the present paper, variance identification is equally important.

\paragraph*{Rotational invariance.}
Let us assume for the moment
that $\facloadtrue  \trans{\facloadtrue }$ is identified.
Consider, for further illustration, the following factor loading matrix
$\facloadtrue $ and a  loading matrix $\facload =  \facloadtrue   \Pm_{\alpha b} $ defined as
a rotation of $\facloadtrue $:
\begin{eqnarray}  \label{example1}
\facloadtrue  = \left(\begin{array}{cc}
	\lambda_{11} & 0 \\
	\lambda _{21} & 0 \\
	\lambda_{31} & 0 \\
	0 & \lambda_{42} \\
	0 & \lambda_{52}\\
	0 & \lambda_{62}
\end{array}\right), \quad \Pm_{\alpha b} = \left(
\begin{array}{rr}
	\cos \alpha & (-1)^b\sin \alpha \\
	-\sin \alpha & (-1)^b\cos \alpha \\
\end{array}
\right),
\quad
\facload= \facloadtrue \Pm_{\alpha b} =  \left(\begin{array}{cc}
	\beta_{11} & \beta_{21} \\
	\beta _{21} & \beta _{22} \\
	\beta_{51} & \beta_{52} \\
	\beta _{41} & \beta_{42} \\
	\beta _{51} & \beta_{52}\\
	\beta _{61} & \beta_{62}
\end{array}\right).
\end{eqnarray}
For any
$\alpha \in [0, 2\pi)$ and $b\in\{0,1\}$,
the factor loading matrix
$\facloadtilde $
yields the same cross-covariance matrix for $\ym_t$ as $\facloadtrue$, as is easily verified:
\begin{eqnarray}
\facloadtilde  \trans{\facloadtilde}  =
\facloadtrue \Pm_{\alpha b} \trans\Pm_{\alpha b}  \trans{\facloadtrue } = \facloadtrue  \trans{\facloadtrue }.
\label{rotation}
\end{eqnarray}
The rotational invariance apparent in (\ref{rotation})  holds
more generally for
any basic factor model (\ref{fac1}). Take any
arbitrary  $\dimmat{\nfactrue}{\nfactrue}$  rotation matrix
$\Pm$  (i.e.\ $\Pm \trans{\Pm}= \identy{\nfactrue}$)
and define  the basic factor  model
\begin{eqnarray}  \label{fac1rot}
\facm_t^{\star}  \sim  \Normult{\nfactrue}{\bfz,\identy{\nfactrue}}, \quad
\ym_t =  \facload \facm^{\star}_t + \errorm_t, \quad \errorm_t
\sim \Normult{\dimy}{\bfz,\Varetrue},
\end{eqnarray}
where $\facloadtilde =  \facloadtrue \Pm$ and
$\facm_t^{\star} =   \trans{\Pm} \facm_t$. Then both models
imply  the
same covariance $\Vary$, given  by (\ref{fac4}).
Hence,  without imposing  further constraints,  $\facloadtrue$ is in general
not identified from  the  cross-covariance matrix $ \facloadtrue  \trans{\facloadtrue} $.
If interest lies in interpreting the factors through the factor loading matrix $\facloadtrue$,
rotational invariance has to be resolved.
The usual way of dealing with rotational invariance is to constrain $\facloadtrue$  in such a
way that the only possible rotation  is the identity $\Pm=\identy{\nfactrue}$.
For orthogonal factors  at least $\nfactrue(\nfactrue-1)/2$ restrictions on the elements of
$\facloadtrue$  are needed to eliminate rotational  indeterminacy \citep{and-rub:sta}.

The most popular constraints are positive lower triangular (PLT) loading matrices, where the upper triangular part is constrained to be zero and the main diagonal elements $\loadtrue_{11},\ldots, \loadtrue_{\nfactrue \nfactrue}$ of $\facloadtrue$ are strictly positive, see Figure~\ref{figglt} for illustration.
Despite its popularity, the PLT structure is  restrictive, as outlined already  by  \citet{joe:gen}.
Let $\facload \trans{\facload} $ be an arbitrary cross-covariance matrix with factor loading matrix $\facload$.  A
PLT  representation  of  $ \facload  \trans{\facload} $
is possible iff a rotation matrix $ \Pm$
exists such that $\facload $ can be rotated into a PLT matrix $\facloadtrue  =  \facload \Pm $.
However, as example (\ref{example1})  illustrates this is not necessarily the case.
Obviously, $ \facloadtrue$ is not a PLT matrix, since
$\loadtrue_{22}=0$. Any of the  possible  rotations
$\facloadtilde =  \facloadtrue \Pm_{\alpha b}$  have {\em non-zero}
elements above the main diagonal and are not PLT matrices either.
This example demonstrates  that  the
PLT  representation is restrictive.
To circumvent this problem in example (\ref{example1}), one could reorder the measurements
in an appropriate manner. However, in applied
factor analysis, such an appropriate ordering is typically not known in advance and
the  choice of the first
$\nfactrue$ measurements
is an important modeling decision under PLT constraints, see e.g.\ \citet{lop-wes:bay} and~\citet{car-etal:hig}.

We discuss in Section~\ref{uniqload} a new  identification strategy to resolve rotational invariance in factor models
based on the concept of generalized lower triangular (GLT) structures.
Loosely speaking, GLT structures generalize PLT structures by
freeing the position of the first non-zero factor loading in each
column, see the loading matrix $ \facloadtrue$ in  (\ref{example1})
and Figure~\ref{figglt} for an example.
We show in Section~\ref{secGLT} that
a unique GLT structure $\facloadtrue$ can be identified for
any cross-covariance matrix $\facload  \trans{\facload}$, provided that
variance identification holds and, consequently,
$\facload  \trans{\facload}$ itself is identified.
Even if   $ \facload  \trans{\facload} $ is obtained
from a loading matrix $\facload$  that does not take the form of a
GLT structure,
such as the matrix $\facload$ in (\ref{example1}),
we show in Section~\ref{secRREF}
that a {\em unique} orthogonal matrix $\rotG$ exists which represents
$\facload$ as a rotation of a unique GLT structure $\facloadtrue$:
\begin{eqnarray} \label{rotRREFint}
\facloadtrue= \facload  \rotG, %
\end{eqnarray}
which we call %
rotation into GLT. Hence,
the GLT  representation is unrestrictive in the sense of \citet{joe:gen}
and  is, indeed, a new and generic way to resolve rotational invariance
for any factor loading matrix.

\paragraph*{Sparse factor loading matrices.}
The factor loading matrix $\facloadtrue$ given in (\ref{example1}) is an
example of a sparse loading matrix. While only a single zero loading would
be needed to resolve rotational invariance,
six zeros are present and each factor loads only  on dedicated
measurements. Such  sparse  loading matrices
are generated by a binary indicator matrix $\deltav$
of 0s and 1s of the same dimension as  $\facloadtrue$,
where  $\loadtrue_{ij}=0$  iff $\delta_{ij}=0$, and
$\loadtrue_{ij} \in \Real$ is unconstrained otherwise.
The binary matrix $\deltav=\indic{\facloadtrue \neq 0}$, where the indicator function
is applied element-wise, is called the sparsity matrix  corresponding to $\facloadtrue$.
The sparsity matrix $\deltav$ contains a lot of information about the structure of
$\facloadtrue$, see Figure~\ref{figglt} for illustration.
The indicator matrix on the right hand side tells us that $\facloadtrue$ obeys
the PLT constraint.
The fifth row of the left and center matrices contains only zeros, which tells us that observation $y_{5t}$ is uncorrelated with the remaining observations,
since $\Cov{y_{it}, y_{5t}} =  0$ for all $i\neq 5$.

\paragraph*{Variance identification.}
Constraints that resolve  rotational invariance
typically take variance identification, i.e.\ identification of
$\facloadtrue  \trans{\facloadtrue }$, for granted, see e.g.\ \citet{gew-zho:mea}.
Variance identification refers  to the problem that
the idiosyncratic variances $\idiov_1, \ldots, \idiov_\dimy$ in $\Varetrue$
are identified  only from the diagonal elements
of $\Vary$, as all other elements are independent of the $\idiov_i$s; see again \eqref{fac5}.
To achieve variance identification of $\idiov_i$ from  $\Omega_{ii}=
\facloadtrue _{i,\bullet} \trans{\facloadtrue _{i,\bullet}} + \idiov_i$,
all factor loadings
have to be identified solely from the off-diagonal elements of $\Vary$.
Variance identification,
however, is easily violated,
as the following considerations illustrate.

Let us return to the factor model defined in (\ref{example1}).
The corresponding covariance  matrix $\Vary$ is given by:
{\begin{eqnarray} \small
	\Vary = \left(
	\begin{array}{cccccc}
		{  \lambda_{11}^2 + {\idiov_1} } & { \lambda_{11}   \lambda_{21}} & { \lambda_{11}  \lambda_{31}}  & && \\
		{ \lambda_{11}  \lambda_{21}} & { \lambda_{21}^2 + {\idiov_2} } & { \lambda_{21}  \lambda_{31}}  & &
		\bfz  &\\
		{ \lambda_{11}   \lambda_{31}} & { \lambda_{21}   \lambda_{31}} & { \lambda_{31}^2 + {\idiov_3} }  & &&  \\
		& &&   \lambda_{42}^2 + {\idiov_4} & \lambda_{42}  \lambda_{52} & \lambda_{42}  \lambda_{62}  \\
		& \bfz  &  & \lambda_{42}  \lambda_{52} & \lambda_{52}^2 + {\idiov_5} & \lambda_{52}  \lambda_{62}    \\
		& & & \lambda_{42}  \lambda_{62} & \lambda_{52}  \lambda_{62}& \lambda_{62}^2 + {\idiov_6} \\
	\end{array}\right).  \label{Omegatrue}
\end{eqnarray}}
Let us assume that the sparsity pattern  $\deltav=\indic{\facloadtrue \neq 0}$
of $\facloadtrue$ is known, but the specific values of the
unconstrained loadings $(\lambda_{11}, \ldots, \lambda_{62})$ are
unknown.
An interesting question is the following. Knowing $\Vary$,
can  the unconstrained loadings $\lambda_{11}, \ldots,
\lambda_{62}$ and the variances $\idiov_1, \ldots, \idiov_\dimy$ be identified uniquely?
Given $\Vary$, the  three nonzero  covariances $\Cov{y_{1t}, y_{2t}}= \lambda_{11}   \lambda_{21}$,
$\Cov{y_{1t}, y_{3t}}=\lambda_{11}   \lambda_{31}$
and $\Cov{y_{2t}, y_{3t}}=\lambda_{21}   \lambda_{31}$  %
are available to identify the three factor loadings $(\lambda_{11}, \lambda_{21}, \lambda_{31})$.
Similarly,
the nonzero  covariances $\Cov{y_{4t}, y_{5t}}= \lambda_{42}   \lambda_{52}$,
$\Cov{y_{4t}, y_{6t}}=\lambda_{42}   \lambda_{62}$
and $\Cov{y_{5t}, y_{6t}}=\lambda_{52}   \lambda_{62}$  %
are available to identify the factor loadings $(\lambda_{42}, \lambda_{52}, \lambda_{62})$,
hence variance identification is given.
However, if we remove the last measurement from the loading
factor matrix defined in (\ref{example1}), we obtain
\begin{eqnarray}  \label{example2}
\facloadtrue  = \left(\begin{array}{cc}
	\lambda_{11} & 0 \\
	\lambda _{21} & 0 \\
	\lambda_{31} & 0 \\
	0 & \lambda_{42} \\
	0 & \lambda_{52}\\
\end{array}\right),
\quad
\facload= \facloadtrue \Pm_{\alpha b}= \left(\begin{array}{cc}
	\beta_{11} & \beta_{21} \\
	\beta _{21} & \beta _{22} \\
	\beta_{51} & \beta_{52} \\
	\beta _{41} & \beta_{42} \\
	\beta _{51} & \beta_{52}\\
\end{array}\right),
\end{eqnarray}
and the corresponding covariance matrix reads:
\begin{eqnarray*} {\small
	\Vary = \left(
	\begin{array}{ccccc}
		{  \lambda_{11}^2 + {\idiov_1} } & { \lambda_{11}   \lambda_{21}} & { \lambda_{11}  \lambda_{31}}  & & \\
		{ \lambda_{11}  \lambda_{21}} & { \lambda_{21}^2 + {\idiov_2} } & { \lambda_{21}  \lambda_{31}}  &
		\bfz  &\\
		{ \lambda_{11}   \lambda_{31}} & { \lambda_{21}   \lambda_{31}} & { \lambda_{31}^2 + {\idiov_3} }  & &  \\
		& &&   \lambda_{42}^2 + {\idiov_4} & \lambda_{42}  \lambda_{52}  \\
		& \bfz  & & \lambda_{42}  \lambda_{52} & \lambda_{52}^2 + {\idiov_5} \large
	\end{array}\right)}.
\end{eqnarray*}
While the three factor loadings $(\lambda_{11}, \lambda_{21}, \lambda_{31})$ are still identified from
the off-diagonal elements of $\Vary$ as before,
variance identification of $\idiov_4$ and $\idiov_5$ fails.
Since $\Cov{y_{4t}, y_{5t}}= \lambda_{42}   \lambda_{52}$ is the only non-zero
element %
that depends on
the loadings $\lambda_{42}$ and $\lambda_{52}$,
infinitely many different parameters $(\lambda_{42},\lambda_{52},
\idiov_4,\idiov_5)$ imply the same covariance matrix $\Vary$.
From these considerations  it is evident that a minimum of three
non-zero loadings is necessary
in each column to achieve  variance identification, a condition
which has been noted as early as \citet{and-rub:sta}. At the same time,
this condition
is not sufficient, as it is satisfied by the loading matrix $\facload$ in
\eqref{example2}, although variance identification does
not hold.
In general, variance identification is not straightforward
to verify.  We will introduce in Section~\ref{sec:3579} a new and convenient way to
verify variance identification for GLT structures. 

\paragraph*{The row deletion property.}
As explained above, we need to verify
uniqueness of the variance decomposition,  i.e.\ the identification of the idiosyncratic
variances $\idiov_1, \ldots, \idiov_\dimy$  in
$\Varetrue$ from the covariance matrix $\Vary$ given in (\ref{fac4}).
The identification of  $\Varetrue$ guarantees  that   $\facloadtrue  \trans{\facloadtrue }$ is
identified.  %
The second step of identification is then
to ensure  uniqueness of the factor loadings, i.e.\ unique identification of $\facloadtrue$
from $\facloadtrue  \trans{\facloadtrue }$.
To verify variance identification,
we rely in the present paper on  a condition
known as {\em row-deletion property}.

\begin{defn}[Row deletion property \AR\ \citep{and-rub:sta}] \label{ARdef}
An  $\dimy \times \nfactrue$ factor loading matrix $\facloadtrue$
satisfies the row-deletion property if the following condition is satisfied:
whenever an arbitrary row is deleted from $\facloadtrue$, two disjoint submatrices of rank $\nfactrue$ remain.
\end{defn}

\noindent \citet[Theorem~5.1]{and-rub:sta} prove that the row-deletion property
is a sufficient condition
for
the identification of $\facloadtrue  \trans{\facloadtrue} $ and  $\Varetrue$
from the marginal covariance matrix $\Vary$  given in (\ref{fac4}).
For any (not necessarily GLT) factor loading matrix $\facloadtrue$, the
row deletion property \AR\
can be trivially tested by a step-by-step analysis, where each single row of
$\facloadtrue$ is sequentially deleted and the two distinct submatrices are determined
from examining the remaining matrix, as suggested e.g.\ by \citet{hay-mar:exa}.
However, this procedure is inefficient and challenging in higher dimensions.

Hence, it is helpful to have more structural conditions for verifying variance identification
under the row deletion property \AR .
The literature provides several necessary
conditions for the row deletion property \AR\ that are based on
counting the number of non-zero factor loadings in $\facloadtrue$.
\citet{and-rub:sta}, for instance,  prove
the following necessary conditions for  \AR : for every nonsingular   $\nfactrue$-dimensional
square matrix  $\Gm$,
the matrix $\facload=\facloadtrue \Gm$ contains  in each column \emph{at least 3}
and in each pair of columns   \emph{at least 5}   nonzero factor loadings.
\citet[Theorem~3.3]{sat:stu} extends  these {\em necessary}
conditions in the following way: every subset of
$1\leq q \leq  \nfactrue$  columns
of  $\facload=\facloadtrue \Gm$ contains \emph{at least $2q+1$}  nonzero factor loadings for every
nonsingular matrix $\Gm$. We call this the $\CountAR$ counting rule for obvious reasons.

For illustration, let us return to the examples in (\ref{example1}) and (\ref{example2}).
First, apply the $\CountAR$ counting rule to the unrestricted matrix  $\facload$ in (\ref{example2}).
Although the variance decomposition
$\Vary=     \facload  \trans{\facload } + \Vare
= \facloadtrue \trans{\facloadtrue } + \Varetrue,$  %
is not unique, the counting rules are not violated,
since $\facload$  has five non-zero rows
except for the cases $(\alpha,b) \in \{0, \frac{\pi}{2}, \pi, \frac{3\pi}{2}\}\times \{0,1\}$.
Only  for these eight specific cases, which correspond to the  trivial rotations
\begin{eqnarray}  \label{example3}
&&  %
\left(\begin{array}{cc}
	\lambda_{11} & 0 \\
	\lambda _{21} & 0 \\
	\lambda_{31} & 0 \\
	0 & \lambda_{42} \\
	0 & \lambda_{52}\\
\end{array}\right) \quad
\left(\begin{array}{cc}
	\lambda_{11} & 0 \\
	\lambda _{21} & 0 \\
	\lambda_{31} & 0 \\
	0 & -\lambda_{42} \\
	0 & -\lambda_{52}\\
\end{array}\right) \quad
\left(\begin{array}{cc}
	-\lambda_{11} & 0 \\
	-\lambda _{21} & 0 \\
	-\lambda_{31} & 0 \\
	0 & \lambda_{42} \\
	0 & \lambda_{52}\\
\end{array}\right) \quad
\left(\begin{array}{cc}
	-\lambda_{11} & 0 \\
	-\lambda _{21} & 0 \\
	-\lambda_{31} & 0 \\
	0 & -\lambda_{42} \\
	0 & -\lambda_{52}\\
\end{array}\right)\\
&&  \nonumber \left(\begin{array}{cc}
	0 & \lambda_{11}  \\
	0 &  \lambda _{21}  \\
	0 & \lambda_{31}  \\
	\lambda_{42} & 0\\
	\lambda_{52} & 0\\
\end{array}\right) \quad
\left(\begin{array}{cc}
	0 & -\lambda_{11}  \\
	0 & - \lambda _{21}  \\
	0 & -\lambda_{31}  \\
	\lambda_{42} & 0\\
	\lambda_{52} & 0\\
\end{array}\right) \quad
\left(\begin{array}{cc}
	0 & \lambda_{11}  \\
	0 & \lambda _{21}  \\
	0 & \lambda_{31}  \\
	-\lambda_{42} & 0\\
	-\lambda_{52} & 0\\
\end{array}\right) \quad
\left(\begin{array}{cc}
	0 & -\lambda_{11}  \\
	0 & -\lambda _{21}  \\
	0 & -\lambda_{31}  \\
	-\lambda_{42} & 0\\
	-\lambda_{52} & 0\\
\end{array}\right),
\end{eqnarray}
we find immediately that the counting rules are violated, since
one of the two columns  has only two non-zero elements.
This example shows the need to check such counting rules not only
for a single loading matrix $\facload$, but also for all  rotations $\facload \Pm$
admissible under the chosen strategy toward rotational invariance.
On the other hand, if we apply the $\CountAR$ counting rule to
the unrestricted matrix  $\facload$ in (\ref{example1}),
we find  that the {\em necessary} counting rules are satisfied for all rotations $\facload \Pm_{\alpha b}$.
For this specific example,  we have already verified explicitly that variance identification holds
and one might wonder if, in general, the $\CountAR$ counting rule can lead to
a {\em sufficient} criterion for variance identification under \AR .

Sufficient conditions for variance identification are hardly investigated in the literature.
One exception is  the popular factor analysis model where $\facloadtrue$
takes the form of a {\em dense} PLT matrix, where 
all  factor loadings %
on and below the main diagonal are left unrestricted and can take any in value in $\Real$.
For this model, condition \AR\ and hence variance identification
holds, except for a set of measure 0, if the condition
$\dimy \geq 2\nfactrue +1$ is satisfied.
\citet{con-etal:bay_exp} investigate identification of a dedicated factor model,  where equation
(\ref{fac1}) is
combined with correlated (oblique) factors, $\facm_t \sim  \Normult{\nfactrue}{\bfz,\Rm}$,
and the factor loading matrix $\facloadtrue$  has a  perfect simple structure, i.e.\ each observation
loads on at most one factor, as in (\ref{example1}) and (\ref{example2}); however, the exact
position of the non-zero elements is unknown.
They prove
necessary and sufficient conditions  that imply uniqueness of the variance decomposition
as well as uniqueness of the
factor loading matrix, %
namely:
the correlation matrix $\Rm$ is of full rank ($\rank{\Rm}=\nfactrue$) and
each column of $\facloadtrue$ contains at least three nonzero loadings.

In the present paper, we build on and extend this previous work.
We provide
sufficient conditions for variance identification of a GLT structure $\facloadtrue$.
These  conditions  are formulated as counting rules
for the $\dimy \times \nfactrue$
sparsity matrix $\deltav = \indic{\facloadtilde\neq 0}$ of $\facloadtilde$
and are
equivalent to the $\CountAR$ counting rules of \citet[Theorem~3.3]{sat:stu}.
More specifically, if
the  $\CountAR$ counting rule holds  for the sparsity matrix $\deltav$ of a GLT matrix
$\facloadtrue$, then this is a sufficient condition
for the row deletion property \AR\ and consequently for variance identification, except for a set of measure 0.

\paragraph*{Identification of the number of factors.}

Identification of the number of factors is a notoriously difficult problem and
analysing this problem from the view point of variance identification is helpful
in understanding some fundamental difficulties.
Assume that $\Vary$ has a representation as in (\ref{fac4}) with $\nfactrue$ factors
which is variance identified.
Then, on the one hand, no equivalent representation exists with $\nfactrue' < \nfactrue$ number of factors.
On the other hand, as shown in
\citet[Theorem~3.3]{rei:ide}, any such structure $(\facloadtrue,\Varetrue)$
creates  solutions  $(\facload_\nfac, \Vare_\nfac)$ with $\dimy \times \nfac$ loading matrices
$\facload_k$  of dimension  $\nfac= \nfactrue+1, \nfactrue+2, \ldots, \dimy$ bigger than $\nfactrue$
and $\Vare_\nfac$ being a positive definite matrix different from $\Varetrue$
which imply the same covariance matrix $\Vary$ as $(\facloadtrue,\Varetrue)$,
i.e.:
\begin{eqnarray}
\Vary=     \facload_\nfac   \trans{\facload}_\nfac + \Vare_\nfac .   \label{facover}
\end{eqnarray}
Furthermore,  for any  fixed $\nfac > \nfactrue$,
infinitely many such solutions $(\facload_\nfac, \Vare_\nfac) $ can be created
that satisfy the decomposition (\ref{facover}) which,  consequently,
no longer is variance identified.
This problem is prevalent regardless
of the chosen strategy toward rotational invariance.
For illustration, we return to  example (\ref{example1}) and
construct an equivalent solution for  $\nfac=3$.
While the first two columns  of  $\facload_3$ are equal to $\facloadtrue$,
the third column is  a so-called {\em spurious} factor with a single non-zero loading
and  $\Vare_3 $ is defined as follows:
\begin{eqnarray} \label{example4}
\facload_3  = \left(\begin{array}{ccc}
	\lambda_{11} & 0 & 0\\
	\lambda _{21} & 0 & \load_{23} \\
	\lambda_{31} & 0 & 0 \\
	0 & \lambda_{42} & 0 \\
	0 & \lambda_{52} & 0\\
	0 & \lambda_{62}  & 0
\end{array}\right), \quad
\Vare_3 =  \Diag{\idiov_1,\idiov_2 - \load_{23}^2 ,\idiov_3 , \idiov_4,\idiov_5,\idiov_6}.
\end{eqnarray}
We can place the  spurious factor loading $\load_{i3}$ in any row $i$
and $\load_{i3}$  can take any value
satisfying $0 < \load_{i3}^2 < \sigma^2_i$.
It is easy to verify that any such pair
$(\facload_3,\Vare_3)$ indeed implies the same covariance matrix $\Vary$ as in (\ref{Omegatrue}).

This ambiguity in an overfitting model 
renders the estimation of true number of factors 
$\nfactrue$ a challenging problem and 
leads to considerable 
uncertainty how to choose the number of factors in applied factor analysis.
In Section~\ref{secEFA}, we follow up on  this problem in more detail.
An important necessary condition for $k$ to be the true number of factors
is that variance identification of
$\Vare_\nfac$ in (\ref{facover}) holds. Therefore, the
counting rules that we introduce in this paper will also  be useful
in  cases where the true number of factors $\nfactrue$ is unknown.

\paragraph*{Overfitting GLT structures.}
Finally, we investigate in Section~\ref{secEFA} the class of potentially overfitting
GLT structures where the matrix $\facload_\nfac $ in (\ref{facover}) is constrained
to be an unordered GLT structure. We apply results by \citet{tum-sat:ide}
to this class  and show how easily
spurious factors and the underlying true factor loading matrix $\facloadtrue$
are  identified under GLT structures, even if the model is overfitting.  Our strategy relies on
the concept of extended variance identification
and the extended row deletion property introduced by \citet{tum-sat:ide},
where more than one row is deleted from the loading
matrix. An extended counting rule will be introduced
for the sparsity matrix of a GLT loading matrices $\facload_\nfac$ in Section~\ref{varidesp}
which is useful in this context.

\section{Solving rotational invariance through GLT structures} \label{uniqload}

\subsection{Ordered and unordered GLT structures} \label{secGLT}

In this work, we introduce a new  identification strategy to resolve rotational invariance
based on the concept of generalized lower triangular (GLT) structures.
First, we introduce the notion of {\em pivot rows} of a factor loading matrix $\facloadtrue$.

\begin{defn}[\bf Pivot rows]    \label{Pivotdef}
Consider an $\dimmat{\dimy }{\nfactrue}$ factor loading matrix  $\facloadtrue$
with  $\nfactrue$ non-zero columns.
For each column $j=1, \ldots, \nfactrue$ of
$\facloadtrue$,
the pivot row $l_j$ is defined as the row index of the first
non-zero factor loading in column $j$,  i.e.\ $ \loadtrue_{ij}=0, \forall \, i<l_j$ and
$\loadtrue_{l_j,j} \neq 0$.
The factor loading $\loadtrue_{l_j,j}$ is called the leading factor loading of column $j$.
\end{defn}

\noindent
For PLT factor loading matrices the pivot rows lie on the main diagonal,
i.e.\
$(l_1, \ldots, l_\nfactrue)=(1,\ldots,\nfactrue)$,
and the leading factor loadings   $\loadtrue_{jj} > 0$ are  positive
for all columns $j=1,\ldots,\nfactrue$.
GLT structures generalize the PLT  constraint
by freeing the pivot rows  of a factor loading matrix $\facloadtrue$ and
allowing them  to take arbitrary positions $(l_1, \ldots, l_\nfactrue)$, %
the
only constraint being that the pivot rows  are pairwise distinct.
GLT structures  contain  PLT matrices as the special case where $l_j= j$
for $j=1,\ldots,\nfactrue$.
Our generalization is particularly useful if the ordering
of the measurements $y_{it}$ is in conflict with the PLT assumption. Since
$\loadtrue_{jj}$ is allowed to be 0,  measurements different from the first
$\nfactrue$  ones  may lead the  factors. For each factor $j$, the  leading variable is the
response variable $y_{l_j,t}$ corresponding to the pivot row $l_j$.

We will distinguish between two types of GLT structures, namely ordered and unordered GLT structures.
The following definition introduces ordered GLT matrices.
Unordered GLT structures will be motivated and defined  below. %
Examples of ordered and unordered
GLT matrices
are displayed in  Figure~\ref{figglt} for a model
with $\nfactrue=6$ factors.

\begin{defn}[\bf Ordered GLT structures]    \label{GLTdef}
An $\dimmat{\dimy }{\nfactrue}$ factor loading matrix  $\facloadtrue$ with full column rank $\nfactrue$
has an ordered GLT structure if the pivot rows  $l_1 , \ldots , l_\nfactrue$ of $\facloadtrue$ are ordered, i.e.\
$l_1 < \ldots <  l_\nfactrue$, and the leading factor loadings are positive, i.e.\
$\loadtrue_{l_j,j} > 0$ for $j=1,\ldots,\nfactrue$.
\end{defn}

\noindent Evidently,  imposing an ordered GLT structure resolves  rotational invariance
if the pivot rows are known.  For any two ordered  GLT matrices
$\facloadtilde$ and $ \facloadtrue $  with {\em identical} pivot rows
$l_1 , \ldots , l_\nfactrue$,
the identity      $\facloadtilde =  \facloadtrue \Pm$ evidently
holds iff $\Pm=\identy{\nfactrue}$.
In practice, the pivot rows $l_1 , \ldots , l_\nfactrue$ of a GLT structure are unknown and need to be
identified from the marginal covariance matrix $\Vary$ for a given number of factors $\nfactrue$.
Given variance identification, i.e.\ assuming that the cross-covariance matrix
$\facloadtrue \trans{\facloadtrue}$ is identified,
a particularly important issue for the identification of a GLT factor model
is whether $\facloadtrue $  is uniquely identified from $\facloadtrue \trans{\facloadtrue}$
if the pivot rows $l_1 , \ldots , l_\nfactrue$ are {\em unknown}.
Non-trivial   rotations $\facloadtilde =  \facloadtrue \Pm$ of a loading matrix $\facloadtrue$ with pivot rows $l_1 , \ldots , l_\nfactrue$ might exist such that $\facload \trans{\facload} = \facloadtrue \trans{\facloadtrue} $,
while the pivot rows  $\tilde{l}_1 , \ldots ,  \tilde{l}_\nfactrue$ of $\facloadtilde $ are different
from the pivot rows of $\facloadtrue$.  Very assuringly,
Theorem~\ref{theGLT} shows that this is not the case: not only the pivot rows, but the entire loading matrices
$\facloadtrue$  and $\facloadtilde$ are identical, if $\facloadtrue \trans{\facloadtrue} = \facload \trans{\facload}$   (see Appendix~\ref{app:proof} for a proof).

\begin{thm}\label{theGLT}
An ordered  GLT structure   is uniquely identified, provided that uniqueness of  the variance decomposition holds, i.e.:
if     $\facloadtrue$ and $\facloadtilde$ are  GLT matrices, respectively,  with
pivot rows  $l_1 < \ldots <  l_\nfactrue$ and  $\tilde{l}_1 < \ldots <  \tilde{l}_\nfactrue$
that  satisfy
$\facloadtilde  \trans{\facloadtilde}  =
\facloadtrue  \trans{\facloadtrue }$, then  $\facloadtilde = \facloadtrue$ and
consequently $(\tilde{l}_1 , \ldots ,  \tilde{l}_\nfactrue)=(l_1,\ldots , l_\nfactrue)$.
\end{thm}

Definition~\ref{UGLTdef} introduces, as an extension of Definition~\ref{GLTdef},
unordered GLT structures
under which $\facloadtrue$ %
is identified from $\facloadtrue  \trans{\facloadtrue }$ only up to signed permutations.
A signed permutation  permutes the columns of the factor loading matrix $\facloadtrue$ and switches the
sign of all factor loadings in any specific column.
This leads to a trivial case of  rotational
invariance.
For $r=2$, for instance, the eight signed permutations of the loading matrix $\facloadtrue $ defined
in (\ref{example2}) are depicted in  (\ref{example3}).
More formally, $ \facloadtilde $ is
a  signed permutation of $ \facloadtrue $, iff
\begin{align}\label{eq:Ralpha}
\facloadtilde   =      \facloadtrue  \bP _{\pm} \bP _{\rho} ,
\end{align}
where the permutation matrix $\bP_{\rho}$ corresponds to one of   the $\nfactrue$! permutations  of the $\nfactrue$ columns of $\facloadtrue$
and the reflection matrix
$\bP_{\pm}=\Diag{\pm 1, \ldots, \pm 1}$   corresponds
to one of  the $2^\nfactrue$ ways to  switch the signs of the  $\nfactrue$  columns of $\facloadtrue$.
Often, it is convenient to employ  identification rules
that guarantee identification of $\facloadtrue $
only up to such column and sign switching, %
see e.g.\ \citet{con-etal:bay_exp}.  Any  structure $\facloadtrue$ obeying such an
identification rule represents a whole equivalence class of matrices
given by all  $2^\nfactrue \nfactrue !$ signed permutation
$\facloadtilde   =  \facloadtrue  \bP _{\pm} \bP _{\rho}$ of $\facloadtrue$.
This trivial form of the rotational invariance does not
impose any additional mathematical challenges
and is often  convenient from a computational viewpoint,
in particular for Bayesian inference,
see for e.g.\ \citet{con-etal:bay_exp} and \citet{fru-etal:spa}.

It is easy to verify how identification up to trivial rotational invariance can
be achieved for GLT structures
and
motivates the following definition of unordered  GLT structures
as   loadings matrices $ \facloadtilde$ where
the  pivot rows  $l_1, \ldots , l_\nfactrue$ simply occupy $\nfactrue$ different
rows.  In Definition~\ref{UGLTdef},  no order constraint is imposed
on the pivot rows and no sign constraint is imposed on the leading
factor loadings. This very general structure allows to
design highly efficient sampling schemes for sparse Bayesian factor analysis under GLT structures,
see \cite{fru-etal:spa}.

\begin{defn}[\bf Unordered GLT structures]    \label{UGLTdef}
An $\dimmat{\dimy }{\nfactrue}$ factor loading matrix  $\facload$ with full column rank $\nfactrue$
has an unordered GLT structure if the pivot rows $l_1, \ldots , l_\nfactrue$ of $\facload$
are pairwise distinct.
\end{defn}

\noindent Theorem~\ref{theGLT} is easily  extended to unordered GLT structures.
Any signed permutation $  \facloadtilde =
\facloadtrue \bP _{\rho} \bP _{\pm}$ of $\facloadtrue $
is uniquely identified from $\facloadtilde  \trans{\facloadtilde}= \facloadtrue  \trans{\facloadtrue }$, provided that $\facloadtrue  \trans{\facloadtrue }$ is identified.
Hence, under unordered GLT structures the factor loading matrix $\facloadtrue$
is uniquely identified up to signed permutations.
Full identification can easily be obtained from  unordered GLT structures $\facload$.
Any unordered GLT structure $\facloadtilde$  has unordered pivot rows
$l_1 , \ldots , l_{\nfactrue}$, occupying different rows.
The corresponding ordered GLT structure $\facloadtrue $   is recovered  from  $\facloadtilde$
by sorting the columns in ascending order according to the pivot rows.
In other words,  the pivot rows of $\facloadtrue $ are equal to the order statistics
$l_{(1)} , \ldots ,  l_{(\nfactrue)}$ of
the  pivot rows $l_1, \ldots , l_ \nfactrue$ of $\facloadtilde$,
see again Figure~\ref{figglt}.
This procedure resolves rotational invariance, since the pivot rows
$l_1, \ldots , l_ \nfactrue$ in the unordered GLT structure are distinct.
Furthermore, imposing the  condition  $\loadtrue_{l_j,j}>0$ in each column $j$
resolves sign switching:
if  $\loadtrue_{l_j,j}<0$, then
the sign of all factor loadings  $\loadtrue_{ij}$ in  column $j$ is switched.

\subsection{Sparse GLT structures}  \label{sec:UGLT}

In Definition~\ref{GLTdef} and \ref{UGLTdef},  \lq\lq structural\rq\rq\ zeros are introduced
for a GLT structure for all
factor loading above the pivot row $l_j$, while the factor loading $\loadtrue_{l_j,j}$ in the
pivot row is non-zero by definition.  We call  $\facloadtrue$ a {\em dense} GLT structure
if all loadings
below the pivot row are unconstrained and
can take any value in  $\Real$.

A \emph{sparse} GLT structure results if  factor loadings
at unspecified places below the pivot rows
are zero and only the remaining
loadings  are unconstrained.
A sparse loading matrix $\facloadtrue$ can be characterized by
the so-called sparsity matrix, defined as  a
binary indicator matrix
$\deltav$ of 0/1s of the same size
as $\facloadtrue$,  where $\delta_{ij}=\indic{\loadtrue_{ij} \neq 0}$.
Let $\deltav ^{\Lambda}$ be
the sparsity matrix  of a GLT matrix $\facloadtrue$.
The sparsity matrix  $ \deltav$ corresponding to the signed permutation
$\facloadtilde =
\facloadtrue \bP _{\rho} \bP _{\pm}$ is equal to
$ \deltav = \deltav^{\Lambda} \bP _{\rho}$ and is
invariant to sign switching.
Hence, for any sparse unordered GLT matrix $\facload$,
the corresponding sparsity matrix $\deltav$ obeys an unordered
GLT structure with the same pivot rows as $\facload$,
see Figure~\ref{figglt}  for illustration.

In sparse  factor analysis,
single factor loadings take zero-values with positive probability and
the corresponding sparsity matrix $\deltav$ is a  binary matrix that has to be identified
from the data.
Identification in sparse  factor analysis
has to provide conditions under which
the entire 0/1 pattern in $\deltav$ can be identified from  the
covariance matrix $\Vary$ if  $\deltav$ is unknown.
Whether this is possible hinges on variance identification, i.e.\ whether
the decomposition of  $\Vary$ into $\facloadtrue \trans{\facloadtrue}$  and $\Varetrue$ is unique.
How variance identification can be verified for (sparse) GLT structures is
investigated in  detail in Section~\ref{varidesp}.
Let us assume at this point that variance identification holds, i.e.
the cross-covariance matrix $\facloadtrue \trans{\facloadtrue}$ is identified.
Then an important step toward the identification of a sparse factor model
is to verify whether the 0/1 pattern  of $ \facloadtrue$, characterized by $\deltav$,
is uniquely identified from $\facloadtrue \trans{\facloadtrue}$.
Very importantly, if $\facloadtrue $ is assumed to be
a GLT structure,  then the entire GLT structure $\facloadtrue$ and hence
the indicator matrix  $\deltav$ is uniquely identified from $\facloadtrue  \trans{\facloadtrue }$,
as follows immediately  from Theorem~\ref{theGLT},
since $\delta_{i j}= 0$, iff $\loadtrue_{i j}= 0 $ for all $i,j$.
By identifying  the 0/1 pattern in $\deltav$ we can uniquely  identify  the pivot rows of
$\facloadtrue$ and the sparsity
pattern below.

We would like to emphasize that in sparse factor analysis with unconstrained loading matrices
$\facloadtrue$ this is not necessarily the case. The indicator matrix  $\deltav$ is
in general {\em not} uniquely identified
from $\facloadtrue \trans{\facloadtrue}$, because
(non-trivial)  rotations $\Pm$ %
change  the zero pattern in
$\facloadtilde =  \facloadtrue \Pm$, while $\facloadtilde  \trans{\facloadtilde}  =
\facloadtrue  \trans{\facloadtrue }$.
For illustration, let us return to the example in (\ref{example1}) where  we showed that
$\facloadtrue  \trans{\facloadtrue}$ is uniquely identified if the true
sparsity matrix $\deltav ^{\Lambda}$ is known. Now assume
that $\deltav ^{\Lambda}$ is unknown
and allow the loading matrix
$\facloadtilde =  \facloadtrue \Pm$ to be any rotation of $\facloadtrue$.
It is then evident that the corresponding sparsity matrix $\deltav$ is not unique
and two solutions exists. For all rotations where
$(\alpha,b) \in \{0, \frac{\pi}{2}, \pi, \frac{3\pi}{2}\}\times\{0,1\}$,
$\facloadtilde$ correspond to one of the eight signed permutation of
$\facloadtrue$
given in
\eqref{example3} and the sparsity matrix $\deltav$
is equal to $\deltav^{\facloadtrue}$ up to this signed permutation.
For all other rotations, all elements of $\facloadtilde$ are different
from zero and $\deltav$ is simply a matrix of ones.

\subsection{Rotation into  GLT} \label{secRREF}

As discussed above, GLT structures generalize the PLT constraint, but
one might wonder how restrictive %
this structure 
still is. We will show in this section that for a basic factor model with unconstrained loading matrix
$\facload$ there exists an  equivalent representation involving a unique GLT structure
$\facloadtrue$ which is related to $\facload$  by an orthogonal transformation,
provided that uniqueness of  the variance decomposition holds.

The proof of this %
result uses a relationship between a matrix with GLT structure and the so-called
reduced row  echelon form in linear algebra that results from the Gauss-Jordan elimination
for solving linear systems, see e.g.\ \citet{ant-ror:ele}.  %
Any
transposed GLT
loading matrix $\trans{\facloadtrue}$ has a row echelon form which can be turned  into
a reduced  row echelon form (RREF) $\facloadrref =   \trans{\Am} \trans{\facloadtrue} $
with the help of an
$\dimmat{\nfactrue}{\nfactrue}$ matrix $\Am$ which is constructed from the pivot rows
$l_1,\ldots, l_\nfactrue$ of  $\facloadtrue$ and %
invertible by definition:
\begin{eqnarray*}
\Am ^ {-1} = \left(
\begin{array}{c}
	\facloadtrue_{l_1,\cdot} \\
	\vdots \\
	\facloadtrue_{l_\nfactrue,\cdot}
\end{array}
\right).
\end{eqnarray*}

Since the RREF of any matrix is unique, see e.g.\ \citet{yus:red}, we find that the pivot columns of
$\facloadrref $ coincide with the pivot rows  $l_1,\ldots, l_\nfactrue$ of  $\facloadtrue$.
Hence, for a basic factor model
\begin{eqnarray*} %
\facm_t  \sim  \Normult{\nfactrue}{\bfz,\identy{\nfactrue}}, \qquad  \ym_t =  \facload  \facm_t + \errorm_t,
\end{eqnarray*}
with  an arbitrary, unstructured loading matrix  $\facload$ with full column rank $\nfactrue$,
we prove in Theorem~\ref{therref} that the RREF %
of $\trans{\facload}$ can be used to represent
$\facload$ as a unique GLT structure $\facloadtrue$, where
the pivot rows  $l_1,\ldots, l_\nfactrue$ of  $\facloadtrue$
coincide with  the pivot columns of the RREF of $\trans{\facload}$
(see Appendix~\ref{app:proof} for a proof).

\begin{thm}[Rotation into GLT]\label{therref}
Let  $\facload$ be an arbitrary loading matrix  with full column rank $\nfactrue$. Then the following
holds:
\begin{itemize}
	\item[(a)] There exists an equivalent representation of $\facload$ involving a
	unique GLT structure $\facloadtrue$, %
	\begin{eqnarray} \label{vardecRREF}
		\facload  =  \facloadtrue  \trans{\rotG},
	\end{eqnarray}
	where  $\rotG$ is a unique orthogonal matrix.
	$\facloadtrue$ is called
	the {\em GLT representation} of %
	$\facload$.
	
	\item[(b)]
	Let  $l_1 < \ldots <  l_\nfactrue$ be the pivot columns of the RREF $\facloadrref $
	of $\trans{\facload}$ and let
	$\facload_1$ be the
	$\nfactrue \times \nfactrue$ submatrix
	of $\facload$ containing the
	corresponding rows $l_1, \ldots,  l_\nfactrue$.
	The GLT representation $\facloadtrue = \facload  \rotG$
	of $\facload$  has  pivot rows $l_1 , \ldots ,  l_\nfactrue$ and is obtained
	through  rotation into GLT
	with a rotation matrix
	\begin{eqnarray} \label{vardectilde}
		\rotG = \Qm ,
	\end{eqnarray}
	which results from the QR decomposition  $\Qm \Rm =\trans{\facload_1}$ of $\trans{\facload_1}$.
\end{itemize}
\end{thm}

Would it be possible to obtain a similar
results %
with the factor loading matrix $\facloadtrue$  being constrained to be a  PLT structure?
The answer is definitely no,
as has already been established in Section~\ref{secmotivate} for example (\ref{example1}).
As mentioned above, GLT
structures encompass PLT structures as a special case. Hence,
if a PLT representation $\facloadtrue$
exists for a loading matrix $\facload =\facloadtrue \Pm $,  then the  GLT representation
in (\ref{vardectilde})  {\em automatically} reduces to
the PLT structure $\facloadtrue$, since
$\Rm = \trans{\facload_1}$ is obtained from the first $\nfactrue$ rows of $\facload$ and
the \lq\lq rotation into GLT\rq\rq\
is equal to the identity, $\Qm=\identy{\nfactrue}$.
On the other hand, if  the GLT representation $\facloadtrue$
differs from a PLT structure, then no equivalent
PLT representation exists. Hence, forcing a PLT structure
in the representation (\ref{fac1}) may introduce a systematic bias in estimating the marginal covariance matrix $\Vary$.

\section{Variance identification and GLT structures} \label{varidesp}

As mentioned in the previous sections, constraints imposed on the structure of a factor loading matrix $\facloadtrue$  will resolve rotational invariance  only if uniqueness of the variance
decomposition holds and the cross-covariance matrix $\facloadtrue  \trans{\facloadtrue }$
is identified. However, rotational constraints alone do not necessarily guarantee
uniqueness of the variance decomposition.
Consider, for instance, a sparse PLT  loading matrix where in  some column $j$ in addition
to the diagonal element $\loadtrue_{jj}$ (which is nonzero by definition)  only
a single further factor loading  $\loadtrue_{n_j,j}$ in some row $n_j>j$
is nonzero.
Such a loading matrix obviously violates the necessary  condition for variance identification that
each column contains at least three nonzero elements.
Similarly, while GLT structures resolve rotational invariance, they do not
guarantee  uniqueness  of the variance decomposition either.

In Section~\ref{sec:3579}, we derive sufficient conditions for variance identification
of GLT structures based on the \CountAR\ counting rule
of \citet[Theorem~3.3]{sat:stu}.
In Section~\ref{varbeyGLT}, we discuss how to  verify variance 
identification for sparse GLT structures in practice.

\subsection{Counting rules for variance identification} \label{sec:3579}

We will show how to verify from the 0/1 pattern $\deltav$ of an unordered GLT structure
$\facloadtilde $,
whether the row deletion property \AR\  holds for
$\facloadtilde $ and all its signed permutations. Our condition is
a structural counting rule expressed
solely in terms of
the sparsity matrix $\deltav$ underlying $\facloadtilde $
and does not involve the values of the unconstrained
factor loadings in  $\facloadtilde$, which can take any value in $\Real$.
For any factor model, variance identification is invariant to signed permutations. If
we can verify
variance identification for a single signed permutation
$ \facloadtilde   = \facloadtrue  \bP _{\pm} \bP _{\rho} $
of $\facloadtrue$, as defined in (\ref{eq:Ralpha}), then variance identification
of $\facloadtrue$ holds, since $ \facloadtilde$ and  $\facloadtrue $ imply the same
cross-covariance matrix $\facloadtrue  \trans{\facloadtrue }$.
Hence, we focus in this section on variance identification of unordered GLT structures.

\noindent In Definition~\ref{ARdefext}, we recall the
so-called {\em extended row deletion property}, introduced by
\citet{tum-sat:ide}.

\begin{defn}[Extended row deletion property $\RD{\nfactrue}{\Rs}$] \label{ARdefext}
A  $\dimy \times \nfactrue$ factor loading matrix $\facload$ satisfies the row-deletion property
$\RD{\nfactrue}{\Rs}$, if the following condition is satisfied: whenever
$\Rs \in \mathbb{N}_0$ %
rows are deleted from  $\facload$,
then two disjoint submatrices of rank $\nfactrue$ remain.
\end{defn}

\noindent
The row-deletion property  of \citet{and-rub:sta}
results as a special case where $\Rs={1}$.
As will be shown in Section~\ref{secEFA},
the  extended row deletion properties $\RD{\nfactrue}{\Rs}$ for $\Rs >{1}$ are
useful in exploratory factor analysis,
when the factor dimension $\nfactrue$ is unknown. %
In Definition~\ref{CRdef}, we  introduce a counting rule for binary matrices.

\begin{defn}[Counting rule $\CR{\nfactrue}{\Rs}$] \label{CRdef}
Let $\deltav$  be an $\dimy \times \nfactrue$ binary matrix.
For each $q =1,\ldots,\nfactrue$, consider all submatrices $\deltav_{q,\ell}$, $\ell=1, \ldots, \Bincoeftext{\nfactrue}{q}$,
built from $q$ columns of $\deltav$.
$\deltav$ is said to satisfy the
$\CR{\nfactrue}{\Rs}$ counting rule for $\Rs \in \mathbb{N}_0$
if the matrix
$\deltav_{q,\ell}$ has  at least $2\ell+\Rs $  nonzero rows for all $(q,\ell)$.
\end{defn}

\noindent
Note that the
counting rule $\CR{\nfactrue}{\Rs}$, like the extended row deletion property
$\RD{\nfactrue}{\Rs}$, is invariant to
signed permutations. Lemma~\ref{lemCRrules} in Appendix~\ref{app:proof} summarizes
further useful properties of $\CR{\nfactrue}{\Rs}$.

For a given binary matrix $\deltav$ of dimension $\dimy \times \nfactrue$,
let $\ThetaD{\delta}$ be the space generated by
the non-zero elements of
all unordered  GLT structure
$\facloadtilde$ with sparsity matrix  $\deltav$
and all their $2^\nfactrue \nfactrue ! -1 $ trivial rotations
$\facloadtilde  \bP _{\pm} \bP _{\rho}$. 
We prove in Theorem~\ref{Crule357} that for GLT structures   the counting rule
$\CR{\nfactrue}{\Rs}$  and the extended row deletion property $\RD{\nfactrue}{\Rs}$
are equivalent conditions %
for all loading matrices in  $\ThetaD{\delta}$,  except for a set of measure 0.

\begin{thm}\label{Crule357}
Let $\deltav$  be a binary $\dimy \times \nfactrue$ matrix with unordered GLT structure.
Then the following holds:
\begin{enumerate}
\item[(a)]  If  $\deltav$ violates the counting rule $ \CR{\nfactrue}{\Rs}$,
then   the extended row deletion property $\RD{\nfactrue}{\Rs}$ is violated
for all  $\facloadtilde \in \ThetaD{\delta}$  generated by $\deltav$.
\item[(b)] If $\deltav$ satisfies the counting rule $ \CR{\nfactrue}{\Rs}$,
then the extended row deletion property $\RD{\nfactrue}{\Rs}$  holds %
for all $\facloadtilde \in \ThetaD{\delta}$
except for a set of measure 0.
\end{enumerate}
\end{thm}

\noindent See Appendix~\ref{app:proof} for a proof. The special case
$\Rs=1$ is relevant for verifying the  row deletion property \AR .
It proves   that  for unordered GLT structures
the \CountAR\ counting rule of \citet{sat:stu} is not only a necessary,
but also a {\em sufficient} condition  for \AR\ to hold.
In addition, this means
that the  counting rule needs to be verified
only  for the sparsity matrix $\deltav$  of  a
{\em single  trivial rotation} $ \facloadtilde =\facloadtrue \bP _{\pm} \bP _{\rho}$
rather than for every nonsingular matrix $\Gm$.
This result is summarized in Corollary~\ref{rule357}.

\begin{cor}[Variance identification rule for GLT structures]\label{rule357}
For  any  unordered $\dimy \times \nfactrue$ GLT structure $\facloadtilde $,
the following holds:
\begin{enumerate}
\item[(a)] If $\deltav$ satisfies the
\CountAR\ counting rule,  i.e.\ every column of $\deltav$ has at least 3 non-zero elements,
every  pair of columns at least 5  and, more generally,
every  possible combination of $q=3, \ldots, \nfactrue $ columns has at least $2q+1$ non-zero elements,
then variance identification is given  for all $\facloadtilde \in \ThetaD{\delta}$
except for a set of measure 0;
i.e.\  for any other factor decomposition of the marginal covariance matrix $\Vary =
\facloadtilde \trans{\facloadtilde} + \Varetilde
=     \tilde{\facloadtilde} \trans{\tilde{\facloadtilde}} + \tilde{\Varetilde}$,
where $\tilde{\facloadtilde}$ is an unordered GLT matrix,
it follows  that $\tilde{\Varetilde}= \Varetilde$, i.e.
$\tilde{\facloadtilde} \trans{\tilde{\facloadtilde}}=\facloadtilde \trans{\facloadtilde}$,
and   $\tilde{\facloadtilde} = \facloadtilde \bP _{\pm} \bP _{\rho}$.

\item[(b)] If $\deltav$ violates the
\CountAR\ counting rule, then for all $\facloadtilde \in \ThetaD{\delta}$
the row deletion property \AR\ does not hold.

\item[(c)] For $\nfactrue=1$, $\nfactrue=2$, and $\nfactrue=3$,
condition $\CR{\nfactrue}{1}$ is both sufficient and necessary for variance identification.
\end{enumerate}
\end{cor}

\noindent A few comments are in order.
If $\deltav$ satisfies $\CR{\nfactrue}{1}$,
then  $\AR$ holds for all $\facloadtilde \in \ThetaD{\delta}$
and a sufficient condition for variance identification is satisfied.
As shown by \citet{and-rub:sta},
$\AR$ is a {\em necessary} condition for variance identification only for
$\nfactrue=1$ and  $\nfactrue=2$. \citet[Theorem~3]{tum-sat:ide} show the same for
$\nfactrue=3$, provided that $\dimy\geq 7$.
It follows that $\CR{\nfactrue}{1}$ is a {\em necessary  and sufficient} condition for variance identification
for the models summarized in (c). In all other cases, variance identification may hold
for loading matrices $\facloadtilde \in \ThetaD{\delta}$,
even if  $\deltav$ violates $\CR{\nfactrue}{1}$.

The definition of unordered GLT structures given in Section~\ref{uniqload} imposes no constraint on the pivot
rows $l_1, \ldots , l_{\nfactrue}$ beyond the assumption that they are distinct. This flexibility
can lead to GLT structures that can never satisfy the \CountAR\ rule, even if all elements below
the pivot rows are non-zero.  Consider, for instance, a  GLT matrix with
the pivot row in column  $\nfactrue$ being equal to $l_\nfactrue= \dimy -1$.
The loading matrix has at most two nonzero elements in column $\nfactrue$ and violates the necessary
condition for variance identification. %
This example shows that there is an upper bound for the pivot elements beyond which the \CountAR\ rule
can never hold. This insight is formalized in  Definition~\ref{GLTAR}.

\begin{defn}\label{GLTAR} An unordered GLT structure $\facloadtilde$
fulfills  condition \GLTAR\ if the following constraint on the
pivot rows $l_1, \ldots , l_{\nfactrue}$ of $\facloadtilde$ is satisfied,
where $\rankl_j$ is the rank  of $l_j$ in the ordered sequence $ l_{(1)} <
\ldots < l_{(\nfactrue)}$:
\begin{eqnarray} \label{condlj}
l_j \leq  \dimy - 2(\nfactrue - \rankl_j +1).
\end{eqnarray}
\end{defn}

\noindent Evidently,  an ordered GLT structure $\facloadtrue$
fulfills  condition \GLTAR\ if  the
pivot rows $l_1, \ldots , l_{\nfactrue}$ of $\facloadtrue$   satisfy the constraint
$  l_j \leq  \dimy - 2(\nfactrue - j+1)$.
For the special case of a PLT structure where $l_j=j$, this constraint reduces to
$ \dimy \geq 2\nfactrue  +1 $ which is equivalent to
a well-known  upper bound  for the number of factors.
For dense unordered GLT structures with $m$ (non-zero) rows,
condition \GLTAR\ is a sufficient condition for \AR .
For sparse GLT structures
\GLTAR\ is only a necessary condition for  \AR\   and   the \CountAR\ rule
has to be  verified  explicitly, as shown by the example discussed above.
Very conveniently for verifying variance identification in sparse factor analysis based
on GLT structures,  Theorem~\ref{Crule357} and Corollary~\ref{rule357} operate
solely on the sparsity matrix $\deltav$ corresponding to $\facloadtilde$.

\subsection{Variance identification in practice} \label{varbeyGLT}

To verify $\CR{\nfactrue}{\Rs}$ in practice, all submatrices of $q$  columns have to be
 extracted from
the sparsity matrix $\deltav$ to verify if at least $2q+1$ rows of this
submatrix are non-zero. %
For $q=1,2, \nfactrue -1, \nfactrue$, %
this condition  is easily  verified from simple functionals of $\deltav$,
see  Corollary~\ref{Lemma1} which  follows immediately from Theorem~\ref{Crule357}
(see Appendix~\ref{app:proof} for details).

\begin{cor}[{\bf Simple counting rules  for $\CR{\nfactrue}{\Rs}$}]\label{Lemma1}
Let $\deltav$ be a  $\dimy \times  \nfactrue $ unordered GLT sparsity matrix.
The following conditions on  $\deltav$  are necessary for $\CR{\nfactrue}{\Rs}$ to hold:
\begin{eqnarray}
	&\ones_{\nfactrue \times \dimy} \cdot   \deltav +   \trans{\deltav} ( \ones_{ \dimy \times  \nfactrue}
	- \deltav )  \geq   4 + \Rs - 2 \identy{\nfactrue} , &  \label{checkNC2} \\
	&\ones_{1 \times \dimy} \cdot  \indic{ \deltav^\star >0} \geq  2\nfactrue + \Rs, \quad     \deltav^\star=  \deltav \cdot \ones_{\nfactrue \times 1}, &
	\label{checkNC1} \\
	&  \ones_{1 \times \dimy} \cdot   \indic{\deltav^\star >0}  \geq  2(\nfactrue -1) + \Rs , \quad     \deltav^\star= \deltav ( \ones_{ \dimy \times \dimy} -  \identy{\dimy}),
	& \label{checkNC3}
\end{eqnarray}
where the   indicator function $ \indic{\deltav^\star >0}$ is applied element-wise
and $\ones_{ n  \times k}$ denotes a $ n  \times k$ matrix of ones.
For  $\nfactrue \leq 4$, these conditions are also sufficient for
$\CR{\nfactrue}{\Rs}$ to hold for $\deltav$.
\end{cor}

\noindent Using Corollary~\ref{Lemma1} for $\Rs=1$,
one can efficiently  verify, if the \CountAR\
counting rule  and hence the row deletion property \AR\ holds for unordered GLT factor models
with up to $\nfactrue \leq 4$ factors.
 For   models with more than four factors ($\nfactrue > 4$), a more elaborated strategy is needed.
After checking the conditions of Corollary~\ref{Lemma1}, 
$\CR{\nfactrue}{\Rs}$
could be verified for a  given binary matrix $\deltav$   by
iterating over   all  remaining $\footnotesize{\Bincoefsmall{\nfactrue}{q}}$
subsets  of  $q=3, \ldots, \nfactrue-2$ columns of $\deltav$.
While  this %
is a finite task, such a na\"ive approach may need to visit $2^\nfactrue-1$ matrices in order to make a decision and the  combinatorial explosion quickly becomes an issue in practice as $\nfactrue$ increases.
Recent work by \citet{hos-fru:cov}  establishes the applicability of this framework for large models.

\section{Identification in exploratory factor analysis} \label{secEFA}

In this section, we discuss how the concept of GLT structures is helpful
for addressing identification problems in exploratory factor analysis (EFA).
Consider data $\{\ym_1, \ldots,\ym_T\}$
from a multivariate Gaussian distribution,
$\ym_t  \sim \Normult{\dimy}{\bfz,\Vary}$, where an investigator wants
to perform factor analysis
since she expects that the covariances of the
measurements $y_{it}$ are driven by common factors.
In practice,  the number of factors %
is typically unknown
and often it is not obvious,
whether all $\dimy$ measurements in $\ym_t$ are actually correlated.
It is then common to employ  EFA by fitting
a basic factor model to the entire
collection of measurements in $\ym_t$, i.e.\ assuming the model
\begin{eqnarray}    \label{fac1reg} %
\ym_t =  \facload_k  \facm_t + \errorm_t,  \qquad  \errorm_t \sim \Normult{\dimy}{\bfz,\Vare_k} ,
\end{eqnarray}
with an \emph{assumed} number of factors $\nfac$,
a   $\dimmat{\dimy}{\nfac}$ loading matrix $\facload_k$
with elements $\load_{ij}$
and a diagonal matrix $\Vare_k$ with strictly positive entries.
The EFA model (\ref{fac1reg}) is potentially overfitting
in two ways.
First, the true number of factors $\nfactrue$  is
possibly smaller than $\nfac$, i.e.\ $\facload_k$ has too many columns.
Second, some measurements in $\ym_t$ are possibly irrelevant, which means that $\facload_k$
allows for too many non-zero rows.
The goal is then to determine the true number of factors and to identify
irrelevant measurements from the EFA model (\ref{fac1reg}).

We will address identification under the assumption that the data
are generated by a basic factor model with loading matrix $\betatrue$
with $\nfactrue$ factors which implies  the following  %
covariance matrix $\Vary$:
\begin{eqnarray}
\Vary=     \betatrue  \trans{\betatrue} + \Varetrue.   \label{fac4over}
\end{eqnarray}
Instead of  (\ref{fac4over}), for a given $k$, the EFA model (\ref{fac1reg}) yields the alternative
representation of  $\Vary$:
\begin{eqnarray}
\Vary=     \facload_k   \trans{\facload_k} + \Vare_k  . \label{fac4beta}
\end{eqnarray}
The question is then under which conditions
can the true loading matrix $\betatrue$ %
be recovered from (\ref{fac4beta}).
Let us assume for the moment  that
no constraint  that resolves rotational invariance is imposed on
$\betatrue$ or $\facload_k$.  

\paragraph*{\lq\lq Revealing the truth\rq\rq\ in an overfitting EFA model.} %

A  fundamental problem  in factor analysis %
is the following. If the  EFA  model is overfitting, i.e.\ $\nfac > \nfactrue$,
could we nevertheless recover %
the true loading matrix $\betatrue$ directly from
$\facload_k$? We will show how this can be achieved mathematically by combining
the important work by \citet{tum-sat:ide} with the framework of GLT structures.
We have demonstrated in Section~\ref{secmotivate} using example (\ref{example4})
that solutions in an overfitting model can be constructed by adding spurious columns
\citep{rei:ide,gew-sin:int}.
Additional solutions are obtained as  rotations
of such solutions. For instance,
one of the following solutions may result:
\begin{eqnarray*} %
\tilde{\facload}_3  = \left(\begin{array}{ccc}
0  & \lambda_{11} & 0 \\
\load_{23}  & \lambda _{21} & 0  \\
0  & \lambda_{31} & 0 \\
0  & 0 & \lambda_{42}  \\
0  & 0 & \lambda_{52} \\
0  & 0 & \lambda_{62}
\end{array}\right), \quad
\tilde{\facload}_3  = \left(\begin{array}{ccc}
-\lambda_{11} \sin \alpha             & 0 &  \lambda_{11} \cos \alpha \\
\load_{23} \cos \alpha   -\lambda_{21} \sin \alpha  & 0 & \lambda _{21} \cos \alpha \\
-\lambda_{31} \sin \alpha            & 0 &  \lambda_{31} \cos \alpha\\
0  & \lambda_{42} & 0   \\
0   & \lambda_{52} & 0 \\
0   & \lambda_{62} & 0\\
\end{array}\right),
\end{eqnarray*}
both with the same $\Vare_3$ as in (\ref{example4}). The first case is  a signed permutation
of $\facload_{3}$, while the second case
combines a signed permutation of $\facload_{3}$ with a %
rotation
of the spurious and  $\facloadtrue $'s
first column involving $\Pm_{\alpha b}$.
In the first case, despite the rotation,
both the spurious column and the columns of $\facloadtrue $ are clearly visible,
while in the second case the presence of a spurious column
is by no means obvious and the columns of $\facloadtrue $ are disguised.

In general, for an EFA  model %
that is  overfitting by  a single  column, i.e.\ $\nfac  = \nfactrue+1$,
and $\facload_{\nfac}$ is left unconstrained,
infinitely many representations $(\facload_k,\Vare_k)$ with covariance matrix
$\Vary=   \facload _k  \trans{\facload_k} + \Vare_k$ can be constructed in the following way.
Let the first $\nfactrue$ columns of $\facload_k$ be equal to $\betatrue$ and append an extra column to its right.
In this extra column, which will be called a spurious column, add a single non-zero loading
$\load_{l_\nfac, \nfac}$ in any  row
$1 \leq l_\nfac \leq \dimy $  taking any value that satisfies
$0< \load_{l_\nfac, \nfac}^2< \idiov_{l_\nfac}$;
then reduce the idiosyncratic variance in row $l_\nfac$
to $\idiov_{l_\nfac} - \load_{l_\nfac, \nfac}^2$;
and finally apply an arbitrary rotation $\Pm$:
\begin{eqnarray} \label{adsp}
\facload_{\nfac}=  \left(\begin{array}{cc}
\betatrue  & \left|\begin{array}{c}
	0 \\
	{  \load_{l_\nfac,  \nfac}} \\
	0
\end{array} \right.
\end{array} \right) \Pm  , \qquad
\Vare_{\nfac} =  \Diag{\idiov_1,\ldots, \idiov_{l_\nfac} -
{  \load_{l_\nfac, \nfac}^2}, \ldots, \idiov_\dimy}.
\end{eqnarray}
Interesting questions are then the following: under which conditions
is (\ref{adsp}) an exhaustive representation of all possible solutions
$ \facload_{\nfac}$ in an EFA model where the {\em degree of overfitting}
defined as $s=\nfac-\nfactrue$ is equal to one?
How can all solutions $ \facload_{\nfac}$ be represented if $s>1$?

Such identifiability problems in overfitting EFA models have been analyzed
in depth  by  \citet{tum-sat:ide}.
They show that   a stronger condition
than $\RD{\nfactrue}{1}$ is needed for $\betatrue$ in the
underlying variance decomposition  (\ref{fac4over})  to
ensure that only spurious and no additional common factors are added in
the overfitting representation (\ref{fac4beta}).
In addition, \citet{tum-sat:ide}  provide  a
general  representation of the  factor loading matrix $\facload_k $ in overfitting
representation
(\ref{fac4beta}) with $\nfac  > \nfactrue$.

\begin{thm}\emph{\citep[Theorem~1]{tum-sat:ide}}
Suppose that $\Vary$ has a decomposition  as in  (\ref{fac4over})
with $\nfactrue$ factors   and that
for some  $\tumS \in \mathbb{N}$  with
$\dimy \geq 2\nfactrue + \tumS + 1 $
the  extended row deletion property $\RD{\nfactrue}{1+\tumS}$  holds for $\betatrue$.
If $ \Vary $ has another decomposition such that
$\Vary= \facload_k \trans{\facload_k} + \Vare_k$ where
$\facload_k$ is  a $\dimmat{\dimy}{(\nfactrue+s)}$-matrix of rank
$k=\nfactrue+s$ with $ 1 \leq  s \leq S$,  then there exists  an orthogonal matrix
$\Tm_k$ of rank $k$  such that
\begin{eqnarray}  \label{decover}
\facload_k  \Tm_k =  \left(\begin{array}{cc}
	\betatrue  & \Mm_s
\end{array} \right) ,    \qquad   \Vare_k =  \Varetrue -  \Mm_s  \trans{\Mm_s},
\end{eqnarray}
where  the off-diagonal elements of  $\Mm_s \trans{\Mm_s}$ are zero.
\end{thm}

\noindent The $\dimmat{\dimy}{s}$-matrix  $\Mm_s $
is a  so-called \emph{spurious factor loading matrix} that does not
contribute to explaining the covariance in $\ym_t$, since
\begin{eqnarray*}
\facload_k  \trans{\facload_k } + \Vare_k =  \facload_k \Tm_k \trans{\Tm_k}
\trans{\facload_k} + \Vare_k =
\betatrue  \trans{\betatrue} + \Mm_s  \trans{\Mm_s } + (\Varetrue  -  \Mm_s  \trans{\Mm_s} ) =
\betatrue  \trans{\betatrue} + \Varetrue  =\Vary  .   \label{fac4A}
\end{eqnarray*}

\noindent While this theorem is an important result,
without imposing further structure on the factor loading matrix
$\facload_k$ in the EFA model it cannot be applied immediately
to \lq\lq recover the truth\rq\rq ,
as  the separation of $\facload_k$  into the true factor loading matrix $\betatrue$ and
the spurious factor loading matrix  $\Mm_s$  is possible only up to a  rotation
$\Tm_k$ of  $\facload_k$.
However, %
the truth\rq\rq\ in an overfitting EFA model can be recovered, if  \citet[Theorem~1]{tum-sat:ide}
is applied
within the class of unordered GLT structures introduced in this paper.
If we assume that $\facloadtrue$ is a GLT structure
which satisfies the extended row deletion property $\RD{\nfactrue}{1+\tumS}$,
we prove in Theorem~\ref{theoverGLT} the following result.
If $\facload_k$ in an overfitting EFA model is an unordered GLT structure,
then  $\facload_k$ has a representation,
where the rotation  in (\ref{decover})  is a signed
permutation $\Tm_k = \bP _{\pm} \bP _{\rho}$.
Hence, spurious factors in $\facload_k$
are easily spotted and $\facloadtrue$  can be recovered  immediately from
$\facload_k$.  %

\begin{defn}[Unordered spurious GLT structure] \label{SpurM}
A  $\dimy \times s$  unordered GLT factor loading matrix $\Mm^\Lambda_s$
with pivots rows $\{ {n_1}, \ldots, {n_s} \}$ is an
unordered spurious GLT structure if
all columns are spurious columns with a single nonzero loading in the corresponding
pivot row.
\end{defn}

\begin{thm} \label{theoverGLT}
Let  $\facloadtrue $ be a $\dimy \times \nfactrue$ GLT factor loading matrix
with pivot rows $l_{1} < \ldots < l_{\nfactrue}$ which %
obeys the extended
row deletion property $\RD{r}{1+\tumS}$ for some $\tumS \in \mathbb{N}$.
Assume that the $\dimy \times \nfac$ matrix $\facload_k$ in the EFA variance decomposition
$\Vary=     \facload _k  \trans{\facload_k} + \Vare_k$ is of rank $\rank{\facload _k}=
\nfac=\nfactrue + s$, where
$1 \leq s \leq  \tumS $. If $\facload _k$  is  restricted
to be an  unordered GLT matrix,  then
(\ref{decover}) reduces to
\begin{eqnarray*}
\facload_k \bP _{\pm} \bP _{\rho} = \left(\begin{array}{cc}
	\facloadtrue  & \Mm ^{\Lambda}_s
\end{array} \right),  \quad \Vare_k =  \Varetrue -  \Mm ^{\Lambda}_s
\trans {(\Mm ^{\Lambda}_s)},
\end{eqnarray*}
where $\Mm ^{\Lambda}_s$  is a  spurious ordered GLT structure
with pivot rows ${n_1} < \ldots < {n_s}$
which are distinct from the  $\nfactrue$ pivot rows in $\facloadtrue$.
Hence,    $\nfactrue$ columns  of $\facload_k$
are a signed permutation of  the true loading matrix $\facloadtrue$,
while   the remaining $s$ columns of $\facload_k$ are
an \textit{unordered spurious GLT structure} with  pivots ${n_1}, \ldots , {n_s}$.
\end{thm}

\noindent See Appendix~\ref{app:proof} for a proof.

\paragraph*{Identifying irrelevant variables.} \label{irr}

In applied factor analysis, the assumption  that each measurement $y_{it}$ is correlated with
at least one other measurement is too restrictive,
because irrelevant measurements  might be present
that are uncorrelated with all the other measurements.
As argued by \citet{boi-ng:are}, it is useful to identify such variables.
Within the framework of sparse factor analysis, irrelevant variables are identified
in \citet{kau-sch:ide} by exploring the sparsity matrix $\deltav$ of a factor loading
matrix $\betatrue$  with respect to  zero rows.
Since $\Cov{y_{it}, y_{lt}} = 0$ for all $l \neq i$, if
the entire $i$th row %
of $ \betatrue$ is zero (see also (\ref{fac5})),
the presence of $m_0$ irrelevant measurements causes the  corresponding $m_0$ rows of $ \betatrue$
and $\deltav$ to be zero.
As before, we assume that the variance decomposition (\ref{fac4over})
of the underlying basic factor model is variance identified.

Let us first investigate  identification of  the zero rows in $ \betatrue$ and
the corresponding sparsity matrix $\deltav$
for the case that  the assumed and the true number of factors in
the EFA model (\ref{fac1reg}) are identical, i.e.\ $\nfac = \nfactrue$.
Since variance identification  of (\ref{fac4over})
in the underlying model holds,  we obtain that
$\Varetrue=\Vare_\nfactrue$,  $\betatrue  \trans{\betatrue}=\facload_\nfactrue
\trans{\facload_\nfactrue}$ and
$\facload_\nfactrue = \betatrue \Pm $ is a rotation of $\betatrue$.
Therefore, the position of the zero rows both in $\betatrue$ and $\facload_\nfactrue $
are identical and all irrelevant variables can be identified from $\facload_\nfactrue$
or the corresponding sparsity matrix $\deltav$,
regardless
of the strategy toward rotational invariance.

What makes this task challenging  in applied factor analysis is that
in practice only
the total number $\dimy$ of observations is known, whereas
the investigator is ignorant   both about the
number of  factors $\nfactrue$ and the number of irrelevant measurements $m_0$.
In such a situation,  variance identification of  $\Vare_k$
for an EFA model with $\nfac$ assumed factors is easily lost
if  too many irrelevant variables are included in relation to $\nfac$.
These considerations have important implication for exploratory factor analysis. While the investigator can
choose $ \nfac$, she is ignorant about the number of irrelevant variables
and the recovered model might not be variance identified.
For this reason, it is relevant to verify in any case
that the solution $\facload_\nfac$  obtained
from any EFA model satisfies variance identification.

Under \AR\ this means that
the loading matrix of the correlated measurements, i.e.\
the  non-zero rows of $ \betatrue$,  satisfies $\RD{\nfactrue}{1}$.
If variance identification relies on \AR ,
then  a minimum requirement for $\facload_\nfac$ to satisfy $\RD{\nfac}{1}$ is that
$ 2 \nfac   + 1 \leq \dimy-m_0 $. %
If no irrelevant measurement are present, then the well-known upper bound
$   \nfac  \leq \frac{\dimy-1}{2}$ results. %
However, if irrelevant measurements are present, then
there  is a trade-off between $m_0$  and $\nfac$: the more irrelevant measurements
are included, the smaller  the maximum number of assumed factors $\nfac$ has to be.
Hence, the presence of $m_0$ zero rows in $ \betatrue$, while  $ \facload_\nfac$
in the EFA model is allowed to have $\dimy$ potentially non-zero rows requires stronger
conditions for variance identification
than  for an EFA model where the underlying loading matrix
$ \betatrue$ contains only non-zero rows.
More specifically, for a given number $m_0 \in \mathbb{N}$ of irrelevant  measurements,
variance identification
necessitates the more stringent upper bound
$ \nfac  \leq \frac{\dimy-m_0-1}{2} $, where $\dimy-m_0$ is the number of non-zero rows.
On the other hand, for a given number of factors
$\nfac$ in an EFA model, the  maximum number of irrelevant  measurements
that can be included is given by $  m_0 \leq m -(2\nfac + 1)$.

\paragraph*{Identifying the number of factors through an EFA model.}

Let us assume that the variance decomposition (\ref{fac4over})
of the unknown underlying basic factor model is identified. %
As shown by \citet{rei:ide},
the true number of factors  $\nfactrue$ is equal
to the smallest value $\nfac$ that  satisfies (\ref{fac4beta}).
However, in practice, it is not
obvious how to solve this \lq\lq minimization\rq\rq\ problem.
As the following considerations show, verifying variance identification for
$\facload_\nfac$ in an EFA model
can be  helpful in this regard.

If $\nfactrue$ is unknown, then we need
to find a decomposition of $\Vary$ as in
(\ref{fac4beta})  where $ \Vare_\nfac $ is variance identified.
Since the true underlying decomposition (\ref{fac4over}) is variance identified,
any solution where  $ \Vare_\nfac $ is not variance identified can be rejected.
As has been discussed above, any overfitting EFA model,
where $\nfac > \nfactrue$, has infinitely many decompositions of $\Vary$ and therefore is
never variance identified. Hence, if
any solution $ \Vare_\nfac $ of an EFA model with $\nfac$ assumed factors
is not variance identified, then we
can deduce that $\nfac$ is bigger than $\nfactrue$.
On the other hand, if variance identification holds for $ \Vare_\nfac $,
then the decompositions (\ref{fac4over}) and (\ref{fac4beta}) are equivalent and
we can conclude that $\nfactrue=\nfac$, $\Varetrue= \Vare_\nfac$ and
therefore $ \betatrue \trans{\betatrue}=\facload_\nfac   \trans{\facload}_\nfac$.
As a consequence, we can identify the true loading matrix
$ \betatrue=\facload_\nfac \Pm$ from $\facload_\nfac$ mathematically up to a rotation
$\Pm$ \citep[Lemma~5.1]{and-rub:sta}.

This insight shows that  verifying variance identification
is relevant beyond resolving rotational invariance and is essential for recovering the
true number of factors. This has important implications for applied factor analysis.
Most importantly,
the {\em rank} or the {\em number of non-zero columns}
of a factor loading matrix $\facload_\nfac $ recovered
from an EFA model  with  {\em assumed} number $\nfac$ of factors
might  overfit the true number of factors $\nfactrue$,
if
variance identification for $\Vare_k$ is not satisfied and the variance decomposition is not
unique.
Hence, extracting the number of factors from an EFA model
makes only sense in connection with ensuring that variance identification holds.

\section{Illustrative application} \label{secapll}

\subsection{Sparse Bayesian factor analysis} \label{secPBFA}

A common goal of Bayesian factor analysis is to identify the  unknown factor dimension $\nfactrue$  
of  a factor loading matrix from  the overfitting factor model (\ref{fac1reg}) with  potentially $\nfac>\nfactrue$ factors, 
see, among many others, \cite{roc-geo:fas},  \cite{fru-lop:spa}, and \cite{ohn-kim:pos}. %
 Often, spike-and slab priors are employed,
     where  the elements  $\load_{ij}$ of the loading matrix $\facload_k$ 
     apriori are allowed  to be exactly zero with positive
     probability. 
     This is achieved through a prior on the 
     corresponding $\dimmat{\dimy}{k}$
    sparsity matrix $\deltav_k$. In each column $j$, the  indicators
    $\delta_{i j}$  are active apriori  with a column-specific probability $\tau_j$,
    i.e.~$\Prob{\delta_{i j}=1|\tau_j}=\tau_j$ for $i=1, \ldots,m$, where 
     the slab probabilities $\tau_1, \ldots, \tau_k$ arise 
     from    an exchangeable shrinkage prior: %
     \begin{eqnarray} \label{pri2P}
 \tau_j| k \sim  \Betadis{\betaIBP \frac{\alphaIBP}{\nfac},\betaIBP}, \quad j=1, \ldots,k.
 \end{eqnarray}
  If $\betaIBP$ is unknown, then
(\ref{pri2P})  is called a two-parameter-beta (2PB) prior.
 If  $\betaIBP=1$, then 
 (\ref{pri2P})  is called a one-parameter-beta (1PB) prior and takes the form:
    \begin{eqnarray} \label{prialt}
 \tau_j| k \sim  \Betadis{\frac{\alphaIBP}{\nfac},1}, \quad j=1, \ldots,k.
 \end{eqnarray}
 Prior (\ref{prialt}) converges   to the Indian buffet process  prior \citep{teh-etal:sti}
  for $\nfac \rightarrow \infty$. 
 As recently shown by \citet{fru:gen}, prior (\ref{prialt}) 
 has a representation as a cumulative shrinkage process 
 (CUSP) prior \citep{leg-etal:bay}. 
 
  This specification leads to a Dirac-spike-and-slab prior for the factor loadings,  
 \begin{eqnarray} \label{PriorLL}
 & \load_{i j} | \kappa,  \idiov_i, \tau_{j}
\sim   (1- \tau_{j})  \Delta_{0} + \tau_{j}
\Normal{0, \kappa \idiov_i },\\
 & \nonumber \idiov_i \sim \Gammainv{\csigma,\bsigma}, \quad  \kappa \sim \Gammainv{\ckappa,\bkappa}, &
 \end{eqnarray}
 where the columns of the loading matrix are increasingly pulled toward 0 as the column index increases. 
In (\ref{PriorLL}),  a Gaussian slab distribution is assumed with a random global shrinkage parameter $\kappa$, although other slab distributions are possible, see e.g.\ \citet{zha-etal:bay_gro} and
\citet{fru-etal:spa}.

The hyperparameters  $\alphaIBP$ and $\betaIBP$  are instrumental in controlling prior sparsity. Choosing 
 $\alphaIBP=k$ and $\betaIBP=1$ leads to a uniform distribution for $\tau_j$, with 
 the {\em smallest} slab probability $\tau_{(1)}= \min_{j=1,\ldots,k} \tau_j$ also being uniform, while 
the  largest slab probability  $\tau_{(k)}= \max_{j=1,\ldots,k} \tau_j \sim \Betadis{k,1}$, 
 see \cite{fru:gen}. %
 Such a prior is likely to overfit the number of factors, regardless of all other assumptions. A prior with  $\alphaIBP<k$ and $\betaIBP=1$ induces sparsity, since the {\em largest} slab probability  $\tau_{(k)}  \sim \Betadis{\alphaIBP,1} $, while the smallest slab probability
 $\tau_{(1)} \sim \Betadis{\alphaIBP/k,1}$. To control the small probabilities,
 which are important in identifying the true number of factors, $\alphaIBP$ is assumed to be a random parameter  and learnt from the data under  
 the prior $\alpha \sim \Gammad{\aalpha, \balpha}$. 
 $\betaIBP$ controls the prior information in (\ref{pri2P}).   Priors with  $\betaIBP >1$ and $\betaIBP < 1$, respectively, 
decrease and increase the difference between  $\tau_{(1)}$ and $\tau_{(k)}$. Typically, $\betaIBP$ is unknown and is estimated from the data using the prior  
$\betaIBP \sim \Gammad{\abeta, \bbeta}$.

\paragraph*{MCMC estimation.} For a given choice of
hyperparameters, Markov chain Monte Carlo (MCMC) methods are
applied to sample from the posterior distribution
$p(\facload_k, \Vare_k, \deltav_k|\ym)$, given $T$ 
multivariate observations $\ym=(\ym_1, \ldots, \ym_T)$,
see e.g.\ \citet{kau-sch:bay} among many others. 
In \citet{fru-etal:spa}, such a sampler is developed for GLT factor models. To move between factor models of different factor dimension, \citet{fru-etal:spa} exploit Theorem~\ref{theoverGLT} to add and delete spurious columns through a reversible jump MCMC (RJMCMC) sampler.
For each posterior draw $\facload_k$, the active columns $\facload_r$ (i.e.\ all columns with at least 2 non-zero elements) and the corresponding sparsity matrix $\deltav_r$ are determined. If  $\deltav_r$ satisfies the counting rule $\CR{r}{1}$, then $\facload_r$ is a signed permutation of $\facloadtrue$ with the corresponding covariance matrix 
 $\Vare_r = \Vare_k +  \Mm ^{\Lambda}_s
\trans {(\Mm ^{\Lambda}_s)}$, where $\Mm ^{\Lambda}_s$ contains the spurious columns of 
$\facload_k$. These variance identified draws are kept for further inference and the number of columns of $\facload_r$ is considered a posterior draw of the unknown factor dimension $r$.
This algorithm is easily extended to EFA models without any constraints.  

\begin{table}[t!] \caption{%
Sparse Bayesian factor analysis under GLT  and unconstrained structures (EFA) 
under %
a 1PB prior ($\alpha \sim \Gammad{6,2}$) and a 2PB prior ($\alpha \sim \Gammad{6,2},\betaIBP \sim \Gammad{6,6} $).
GLT and EFA-V
use only the variance identified draws ($M_V$ is the percentage of variance identified draws), EFA uses all posterior draws. 
 }\label{tab1}
 \vspace*{2mm}
 {%
 \begin{tabular}{lllcccc}  %
   \hline
  &  & & $M_V$ & \multicolumn{1}{c}{$\hat{r}$}   & \multicolumn{1}{c}{$p(\hat{r}=\rtrue|\ym)$} 
   &         \multicolumn{1}{c}{$\mbox{\rm MSE}_\Omega$}  \\
 Scenario            &       &   Prior  &  Med(QR)   &  Med(QR)  &  Med(QR)   %
            &   Med(QR)  \\
   \hline    Dedic  & GLT   & 1PB  &  97.0 (91.5,98.3) & 5 (5,5) &  0.90 (0.94,0.99) %
   &  0.018 (0.014,0.030)  \\ 
         &  & 2PB    & 97.6 (87.7,98.9)& 5 (5,5) &  0.99 (0.83,1.00) %
         &  0.019 (0.016,0.027)   \\ 
           \cline{2-7} %
         & EFA  &  1PB & -  & 5 (5,6) &  0.66 (0.09,0.79) %
        &  0.020 (0.015,0.026)  \\ 
           &      &2PB  & - & 5 (5,6) &  0.69 (0.36,0.80) %
           &  0.019 (0.014,0.024)\\  
           \cline{2-7} %
                  & EFA-V   &  1PB &  80.3 (49.8,87.0)  & 5 (5,6) &  0.81 (0.17,0.91) %
        &  0.020 (0.015,0.026)  \\ 
           &      &2PB & 82.6 (63.4,87.9) &  5 (5,6) &  0.84 (0.53,0.92) %
           &  0.019 (0.014,0.024)\\
      \hline     
     Block & GLT  & 1PB    & 96.5 (39.4,98.9) & 5 (5,5) &  0.99 (0.28,0.99) %
     &   0.12 (0.08,0.18) \\ 
            &   & 2PB     &  98.7 (61.9,99.4)  &5 (5,5) &  0.99 (0.54,1.00) %
            &  0.10 (0.08,0.14) \\  
                       \cline{2-7} %
     & EFA       & 1PB   &-&   5 (4,5) &  0.78 (0.22,0.88) %
     &   0.14 (0.11,0.20) \\ 
             &   & 2PB  &-& 5 (4,5) &  0.79 (0.08,0.89)  %
             &  0.12 (0.08,0.24)    \\  
             \cline{2-7} %
     & EFA-V       & 1PB   & 87.0 (55.0,91.5) & 5 (4,5) &  0.89 (0.09,0.96) %
     &   0.14 (0.11,0.20)  \\ 
             &   & 2PB  & 85.9 (28.3,90.4) & 5 (4,5) &  0.92 (0.03,0.97) %
             &  0.12 (0.08,0.24) \\ 
  \hline  Dense  & GLT  & 1PB  &  95.7 (84.6,98.6)    &   5 (5,5) &  0.98 (0.92,0.99) %
 &   0.67 (0.44,1.12)  \\ 
               &   & 2PB  & 99.4 (90.8,99.8)  &  5 (5,5) &  0.99 (0.93,1.00) %
               &   0.68 (0.51,1.18) \\ 
              \cline{2-7} %
     & EFA      &  1PB   &-& 5 (5,6)  & 0.76 (0.43,0.85) %
     &   0.54 (0.39,0.76) \\ 
            &     & 2PB   &-&   5 (5,5) &  0.80 (0.66,0.91) %
            &   0.59 (0.43,0.90)  \\ 
           \cline{2-7} %
     & EFA-V       &  1PB   & 84.4 (76.0,90.2) &  5 (5,6) &  0.89 (0.57,0.95) %
     &   0.54 (0.39,0.76)  \\ 
            &     & 2PB   & 89.7 (80.4,93.9)  &  5 (5,5) &  0.93 (0.77,0.98) %
            &   0.59 (0.43,0.90)\\
   \hline
 \end{tabular}}
  {\footnotesize 
Med is the median and  QR are
the 5\% and the 95\% quantile of the various statistics over the 21 simulated data sets.}
\end{table}

\subsection{An illustrative simulation study} \label{illapp}

 For illustration, we perform  a  simulation study  and consider three different data scenarios with 
 $m=30$  and $T=150$. In all three scenarios, $\rtrue=5$ factors are assumed,
 however, the zero/non-zero pattern is quite different. The first setting is a {\em dedicated} factor model,
 where the first 6 variables load on factor 1, the next 6 variables load on factor 2, and so forth, and the final 6 variables
 load on factor 5. A dedicated factor  model has a GLT structure by definition. The second scenario is a {\em block} factor model, 
 where the first 15 observations load only on factor 1 and 2, while the remaining 15 observations only load on factor 3, 4  and 5 and the covariance matrix has a block-diagonal structure. 
 All loadings within a block are non-zero. The third scenario is a {\em  dense} factor loading matrix 
 without any zero loadings and the corresponding GLT representation has a PLT structure. 
  For all three scenarios, non-zero factor loadings are drawn as
  $\lambda_{ij}=(-1)^{b_{ij}} (1 + 0.1 \Normal{0,1})$, where the exponent $b_{ij}$ is a binary variable with $\Prob{b_{ij}=1}=0.2$.
  In all three scenarios,  $\Vare_0=\identm$.
   21 data sets are sampled under these three scenarios  from the Gaussian factor model
(\ref{fac1}). 

\begin{table}[t!] \caption{%
Bayesian factor analysis under GLT  and unconstrained structures (EFA) 
under %
a uniform prior on $\tau_j$.
GLT and EFA-V
use only the variance identified draws ($M_V$ is the percentage of variance identified draws),  EFA uses all posterior draws. 
 }\label{tab2}
\vspace*{2mm}
 {%
 \begin{tabular}{llcccc}  %
   \hline
  &  &  $M_V$ & \multicolumn{1}{c}{$\hat{r}$}   & \multicolumn{1}{c}{$p(\hat{r}=\rtrue|\ym)$} 
   &         \multicolumn{1}{c}{$\mbox{\rm MSE}_\Omega$}  \\
   Scenario         &        &  Med(QR)   &  Med(QR)  &  Med(QR)   &   Med(QR)%
   \\
   \hline  
      Dedic  & GLT     &    50.6 (32.5,62.2)    
       &  6 (5,7) &    0.38 (0.03,0.68) &    0.02
       (0.01,0.03)   \\ 
         & EFA   & -  & 7 (6,8)   &  0.06 (0,0.12)  &  0.02 (0.02,0.03) \\ 
        & EFA-V  &       36.6 (24.4,44.8)    
 & 6 (5,7)   &  0.17 (0,0.44)   &  0.02 (0.02,0.03)   \\ 
    \hline     
   Block & GLT   & 53.3 (29.3,71.3) &    5 (4,6)&     0.62 (0.18,0.85) &     0.11 (0.08,0.17)          \\  
     & EFA      &- & 
               6 (6,7)  &  0.21 (0.00,0.35)    &  0.13 (0.10,0.19)   \\  
       & EFA-V   &     43.3 (17.3,52.2)   &  5 (5,7)    & 0.47 (0.01,0.62)   &  0.13 (0.11,0.19) \\ 
\hline
  Dense  & GLT   &    62.4 (45.8,71.3)    
       & 5 (5,6) &    0.69 (0.05,0.84)     &0.62 (0.44,1.31) \\  
  & EFA       &-& 6 (6,7)   &  0.12 (0.03,0.34)  &  0.52 (0.42,0.74)  \\                  & EFA-V  & 48.1 (30.1,56.7)  & 5 (5,6)  &    0.46 (0.10,0.63)    & 0.52 (0.42,0.73) \\ 
    \hline
 \end{tabular}}
 
{\footnotesize 
Med is the median and  QR are
the 5\% and the 95\% quantile of the various statistics over the 21 simulated data sets.}
\end{table} 

A sparse overfitting factor model is fitted to each simulated data set
with the maximum number of factors $k= 14$ being equal to the upper bound. 
Regarding the structure, we compare a model where the non-zero columns of  $\facload_\nfac$ 
are left unconstrained with a model where a GLT structure is imposed. %
Inference is based on
the Bayesian approach described in Section~\ref{secPBFA}
with two different shrinkage priors on the sparsity matrix $\deltav_k$:
the 1PB prior (\ref{prialt}) with  random hyperparameter $\alpha \sim \Gammad{6,2}$ and
the 2PB prior (\ref{pri2P}) with random hyperparameters $\alpha \sim \Gammad{6,2}$ and 
$ \betaIBP \sim \Gammad{6,6}$.  
MCMC estimation is run for 3000 iterations after a burn-in of 2000 using the 
RJMCMC algorithm of \citet{fru-etal:spa}.

For each 
of the 21 simulated data sets, we evaluate all 12 combinations of data
scenarios, structural constraints (GLT versus unconstrained) and  priors on the sparsity matrix (1PB versus 2PB)
through Monte Carlo estimates of following statistics:
to assess the performance in estimating
the true number $\rtrue$ of factors,
we consider the mode $\hat{\nfactrue}$ of the posterior distribution $p(\nfactrue|\ym) $
and the magnitude of the posterior ordinate $p(\hat{\nfactrue}=\rtrue|\ym)$. To assess the accuracy in
estimating the %
covariance matrix $\Vary$  of the data, we consider the mean squared error
(MSE) defined by
\begin{eqnarray*}
&& \mbox{\rm MSE}_\Omega=\sum_i\sum_{\ell \leq i}
 \Ew{\left(\Vary_{\nfactrue, i\ell}- \Vary_{i\ell}  \right)^2|\ym}/(m(m+1)/2),
\end{eqnarray*}
which accounts both for posterior variance  and bias of %
the estimated  %
covariance matrix $\Vary_r= \facload_r \facload_r ^\top + \Vare_r$ in comparison to the true matrix. %
Table~\ref{tab1} reports, for all 12 combinations the median, the 5\% and the 95\% quantile of these statistics across all simulated data sets. 
For inference under GLT structures,
posterior draws which are not variance identified have been removed. The fraction of variance identified draws is also reported in the table and is in general pretty high. 
As common for sparse Bayesian  factor 
analysis with unstructured loading matrices, the posterior draws are not screened for variance identification and inference is based on all draws. 

Some interesting conclusions can be drawn from Table~\ref{tab1}. First of all, sparse Bayesian factor 
analysis under the GLT constraint successfully recovers
the true number of factors in all three scenarios. For most of the simulated
data sets, 
the posterior ordinate $p(\hat{\nfactrue}=\rtrue|\ym)$
is larger than 0.9. 
Sparse Bayesian factor 
analysis with unstructured loading matrices  
is also quite successful in recovering 
$\rtrue$, but with less confidence. Both over- and underfitting can be observed and the posterior ordinate $p(\hat{\nfactrue}=\rtrue|\ym)$ is much smaller than under
a GLT structure.
For both structures, the 2PB prior yields higher posterior ordinates than the 1PB prior.   

Recently, \citet{hos-fru:cov} proved that the counting
rule $\CR{r}{1}$ can also be applied to verify variance identification for unconstrained loading matrices.  As  is evident from  Table~\ref{tab1},
the fraction of variance identified draws is however, much smaller than under GLT structures.   
Nevertheless, inference w.r.t.\ to the number of factors can be improved also for an unconstrained EFA model by rejecting all draws that do not obey 
the counting rule $\CR{r}{1}$.

It should be emphasized that the ability of Bayesian factor analysis to recover the number of factors from an overfitting model is closely tied to choosing a suitable shrinkage prior on the sparsity matrix $\deltav_k$.
For illustration, we also consider a uniform prior  for $\tau_j$ and report the corresponding statistics in 
Table~\ref{tab2}.
As expected from the considerations in Section~\ref{secPBFA}, considerable overfitting is observed for all simulated data sets, regardless of the chosen structure.

\section{Concluding remarks} \label{secconcluse}

We have given a full and comprehensive mathematical treatment to generalized lower triangular (GLT) structures, a new identification strategy that improves on the popular positive lower triangular (PLT) assumption for factor loadings matrices.
We have proven that GLT retains PLT's good properties: uniqueness and rotational invariance.
At the same time and unlike PLT, GLT exists for any factor loadings matrix; i.e.\ it is not a restrictive assumption.
Furthermore, we have shown that verifying variance identification under GLT structures is simple and is based purely on the zero-nonzero pattern of the factor loadings matrix.
Additionally, we have embedded the GLT model class into exploratory factor analysis with unknown factor dimension and discussed how easily spurious factors and irrelevant variables are recognized in that setup.
At the end, we demonstrated the power of the framework in a simulation study.

\bibliographystyle{chicago}
\bibliography{references}

\newpage

\appendix

\setcounter{equation}{0}
\setcounter{figure}{0}
\setcounter{table}{0}
\renewcommand{\thesection}{\Alph{section}}
\renewcommand{\thetable}{\Alph{section}.\arabic{table}}
\renewcommand{\thefigure}{\Alph{section}.\arabic{figure}}
\renewcommand{\theequation}{\Alph{section}.\arabic{equation}}

\section{Proofs}   \label{app:proof}

\paragraph*{Proof of Theorem~\ref{theGLT}.}

Assume that two pairs $(\facloadtrue ,\Varetrue)$
and $(\facloadtilde ,\Varetilde)$ satisfy  (\ref{fac4}), where both $\facloadtrue $ and
$\facloadtilde$ are GLT matrices with, respectively, pivot rows $l_1 < \ldots <  l_\nfactrue$
and $\tilde{l}_1 < \ldots <  \tilde{l}_\nfactrue$. Uniqueness of  the variance decomposition (\ref{fac4})
implies
\begin{eqnarray} \label{conglt}
\facloadtrue  \trans{\facloadtrue }  = \facloadtilde \trans{\facloadtilde} .
\end{eqnarray}
We need to prove that all columns of $\facloadtrue $ and $\facloadtilde$ are identical.
First, we prove that %
$l_1=\tilde{l}_1$ by contradiction. Assume $\tilde{l}_1 \neq l_1 $. Exploiting the GLT  structure of both matrices,
we obtain from (\ref{conglt}):
\begin{eqnarray} \label{conglt:eq1}
\loadtrue_{l_1,1}^2= \sum_{j=1}^\nfactrue \loadtruetilde_{l_1,j}^2 \neq 0, \qquad
\sum_{j=1}^\nfactrue \loadtrue_{\tilde{l}_1,j}^2=  \loadtruetilde_{\tilde{l}_1,1}^2 \neq 0.
\end{eqnarray}
Assuming $\tilde{l}_1>l_1 $  implies $\loadtruetilde_{l_1,j}=0$ for $j=1, \ldots,\nfactrue$, and
assuming $l_1>\tilde{l}_1$ implies $\loadtrue_{\tilde{l}_1,j}=0$ for $j=1, \ldots,\nfactrue$, and
both results contradict (\ref{conglt:eq1});
hence $l_1=\tilde{l}_1$. By definition, $\loadtruetilde_{l_1,j}=0$ for $j=2, \ldots,\nfactrue$,
and (\ref{conglt:eq1}) implies
$ \loadtruetilde_{l_1,1}^2 = \loadtrue_{l_1,1}^2$, hence
$\loadtruetilde_{l_1,1} = \loadtrue_{l_1,1}$.
For all $i>l_1$ we obtain from (\ref{conglt}) that
$ \Cov{y_{l_1,t}, y_{i t}}= \loadtrue_{l_1,1}  \loadtrue_{i 1}
= \loadtruetilde_{l_1,1}  \loadtruetilde_{i 1} =  \loadtrue_{l_1,1} \loadtruetilde_{i 1}$.
Since $\loadtrue_{l_1,1} \neq 0$, we obtain $\loadtruetilde_{i 1}=  \loadtrue_{i 1}$
for all $i=l_1, \ldots,\dimy$, hence the first
columns of $\facloadtrue $ and $\facloadtilde$ are identical.
We show identity of  the remaining columns by induction. Assume that the first $q-1$
columns of $\facloadtrue $ and $\facloadtilde$ are identical. %
Similarly as above, we prove  $l_q=\tilde{l}_q$ by contradiction. Exploiting the GLT
structure of both matrices,
we obtain from (\ref{conglt}):
\begin{eqnarray*} %
&& \sum_{j=1}^{q-1} \loadtrue_{l_q,j}^2 + \loadtrue_{l_q,q}^2 =
\sum_{j=1}^{q-1} \loadtruetilde_{l_q,j}^2 + \sum_{j=q}^{\nfactrue}  \loadtruetilde_{l_q,q}^2 =
\sum_{j=1}^{q-1} \loadtrue_{l_q,j}^2 + \sum_{j=q}^{\nfactrue}  \loadtruetilde_{l_q,q}^2 ,\\
&& \sum_{j=1}^{q-1} \loadtrue_{\tilde{l}_q,j}^2 + \sum_{j=q}^{\nfactrue} \loadtrue_{\tilde{l}_q,j}^2 =
\sum_{j=1}^{q-1} \loadtruetilde_{\tilde{l}_q,j}^2 + \loadtruetilde_{\tilde{l}_q,q}^2 =
\sum_{j=1}^{q-1} \loadtrue_{\tilde{l}_q,j}^2 + \loadtruetilde_{\tilde{l}_q,q}^2 . \label{conglt:eq2A}
\end{eqnarray*}
Therefore:
\begin{eqnarray} %
\loadtrue_{l_q,q}^2 = \sum_{j=q}^{\nfactrue}  \loadtruetilde_{l_q,q}^2  \neq 0, \qquad
\sum_{j=q}^{\nfactrue} \loadtrue_{\tilde{l}_q,j}^2 =  \loadtruetilde_{\tilde{l}_q,q}^2 \neq 0 . \label{conglt:eq6}
\end{eqnarray}
Assuming $\tilde{l}_q>l_q $  implies $\loadtruetilde_{l_q,j}=0$ for $j=1, \ldots,\nfactrue$,
and %
assuming $l_q>\tilde{l}_q$ implies $\loadtrue_{\tilde{l}_q,j}=0$ for $j=1, \ldots,\nfactrue$ and
both results contradict (\ref{conglt:eq6});
hence $l_q=\tilde{l}_q$. By definition, $\loadtruetilde_{l_q,j}=0$ for $j=q+1, \ldots,\nfactrue$,
and (\ref{conglt:eq6}) implies
$\loadtruetilde_{l_q,q}^2 = \loadtrue_{l_q,q}^2$, therefore
$\loadtruetilde_{l_q,q} = \loadtrue_{l_q,q}$.
For all $i>l_q$ we obtain from (\ref{conglt}):
\begin{eqnarray*} %
\sum_{j=1}^{q-1} \loadtrue_{l_q,j} \loadtrue_{i j}
+ \loadtrue_{l_q,q}  \loadtrue_{i q} = \sum_{j=1}^{q-1} \loadtruetilde_{l_q,j} \loadtruetilde_{i j}
+  \loadtruetilde_{l_q,q}  \loadtruetilde_{i q} =
\sum_{j=1}^{q-1} \loadtrue_{l_q,j} \loadtrue_{i j} +  \loadtrue_{l_q,q}  \loadtruetilde_{i q}   .
\end{eqnarray*}
Since $\loadtrue_{l_q,q} \neq 0$, we obtain  $\loadtruetilde_{i q}=  \loadtrue_{i q}$  %
for all $i=l_q, \ldots,\dimy$, hence the $q$th
column of $\facloadtrue $ and $\facloadtilde$ is identical.

\paragraph{Proof of Theorem~\ref{therref}.}

The first part of the proof shows existence of representation (\ref{vardecRREF})
for an arbitrary factor loading matrix $\facload$  in a basic factor model where
\begin{eqnarray} \label{facuncproof}
\facm_t  \sim  \Normult{\nfactrue}{\bfz,\identy{\nfactrue}}, \qquad
\ym_t =  \facload  \facm_t + \errorm_t.
\end{eqnarray}
Let $\facloadrref $ be the unique RREF of $\trans{\facload}$.
Since
$\trans{\facload}$ has full row rank, the RREF
$ \facloadrref =   \trans{\Am} \trans{\facload} $
is achieved by left multiplication with a square invertible matrix
$\trans{\Am}$ which is unique.
Using $\Am ^{-1}$, we transform  (\ref{facuncproof}) to a model
with correlated factors:
\begin{eqnarray} \label{factra}
\tilde{\facm}_t = \Am ^{-1} \facm_t  \sim  \Normult{\nfactrue}{\bfz,\Qm}, \qquad
\Qm= (\trans{\Am} \Am ) ^{-1}, \qquad
\ym_t  =   \facload \Am  \Am ^{-1} \facm_t + \errorm_t
= \trans{\facloadrref}  \tilde{\facm}_t + \errorm_t.
\end{eqnarray}
Using a decomposition of $\Qm$ such that $\Qm= \Cm \trans{\Cm}$, we transform  model
(\ref{factra}) to a model with uncorrelated factors,
\begin{eqnarray} \nonumber
\facm_t  \sim  \Normult{\nfactrue}{\bfz,\identy{\nfactrue}},  \qquad
\ym_t = \facloadtrue \facm_t + \errorm_t, \label{factraotr}
\end{eqnarray}
where $\facloadtrue= \trans{\facloadrref}  \Cm  $. If we impose the constraint that $\facloadtrue$ is GLT,
then the $\nfactrue$ pivot columns of $\facloadrref $ define the $\nfactrue$ pivot rows of $\facloadtrue$,
since $\facloadrref $ is the RREF of $\trans{\facloadtrue}$.
The sub matrix $\facloadrref_1$ consisting of the $\nfactrue$ pivot columns
of $\facloadrref $  is equal to the identity matrix, $\facloadrref_1=\identy{\nfactrue}$. Denote by
$\facloadtrue_1$ the lower triangular sub matrix  of $\facloadtrue$ containing the
pivot rows $l_1, \ldots,  l_\nfactrue$. Then $\facloadtrue_1 = \trans{\facloadrref}_1 \Cm =  \Cm $ and
it follows that $\Cm$ is a lower triangular matrix. Hence,  $\Cm $ is unique and given by
the Cholesky decomposition of
$\Qm= ( \trans{\Am} \Am) ^{-1} $.
Using $\facloadtrue= \trans{\facloadrref} \Cm  = \facload \Am \Cm$
we find that
$\facloadtrue = \facload  \rotG$, %
where $\rotG =\Am  \Cm$  is an orthogonal matrix since
\begin{eqnarray} \label{orthRREF}
\rotG \trans{\rotG}  =   \Am  \Cm \trans{\Cm} \trans{\Am } =
\Am  \Qm  \trans{\Am } = \Am  ( \trans{\Am} \Am) ^{-1}   \trans{\Am } = \identy{\nfactrue}.
\end{eqnarray}
Therefore,
$\facload = \facloadtrue  \Pm $,
where $\Pm= \rotG ^{-1}= \trans{\rotG} =
\trans{\Cm} \trans{\Am }$ is unique, since both  $\Am$ and $ \Cm$ are unique.
This proves (\ref{vardecRREF}).

The second part of the proof  shows how   $\rotG $ is constructed.
Let $\facloadrref_1$ be the sub matrix  consisting of the $\nfactrue$ pivot columns
of $\facloadrref $ %
and denote by $\facload_1$ the sub matrix  of $\facload$ containing the
corresponding pivot rows $l_1, \ldots,  l_\nfactrue$.
Since $\facloadrref_1=\identy{ \nfactrue}$, it follows
that
$  \facloadrref_1 = \trans{\Am} \trans{\facload_1} =  \trans{(\facload_1 \Am)}=\identy{ \nfactrue}
$,
and we obtain following simple relationship between $\Am$ and $\facload_1$:
\begin{eqnarray*}
	\Am  = \facload_1 ^{-1}= \left(
	\begin{array}{c}
		\facload_{l_1,\cdot} \\
		\vdots \\
		\facload_{l_\nfactrue,\cdot}
	\end{array}
	\right) ^ {-1}.
\end{eqnarray*}
$\Cm$ is  the lower triangular Cholesky factor in the Cholesky decomposition
of $ \Sm =
\facload_1  \trans{\facload_1}$, since
$\Sm = ( \trans{\Am} \Am) ^{-1} = \Am ^{-1} (\trans{\Am}) ^{-1} = \facload_1  \trans{\facload_1}$.
At the same time, we obtain from the QR decomposition $\Qm \Rm =\trans{\facload_1}$ of  $\trans{\facload_1}$
that
\begin{eqnarray*}
	\facload_1 \trans{\facload_1}= \trans{\Rm} \trans{\Qm}  \Qm \Rm = \trans{\Rm} \Rm.
\end{eqnarray*}
Hence $\Cm= \trans{\Rm}$, since the  Cholesky decomposition is unique, and it follows
that
\begin{eqnarray*}
	\rotG  =   \Am  \Cm = \facload_1 ^{-1} \Rm =   \Qm (\trans{\Rm}) ^{-1}  \trans{\Rm}= \Qm.
\end{eqnarray*}
This proves part~(b).

\paragraph{Properties of the
	counting rules $\CR{\nfactrue}{\Rs}$.} Some useful properties of $\CR{\nfactrue}{\Rs}$ are
summarized in Lemma~\ref{lemCRrules}.  The proof is straightforward.

\begin{lem}\label{lemCRrules}
	The counting rule $\CR{\nfactrue}{\Rs}$ has the following properties:
	\begin{itemize}
   \item[(a)] $\CR{\nfactrue}{\Rs}$  holds for $\deltav$ iff $\CR{q}{s}$
   holds for every submatrix
   of $q \in \{1, \ldots, \nfactrue\}$ columns of $\deltav$.
   \item[(b)] If $\CR{\nfactrue}{\Rs}$ holds for $\deltav$ and arbitrary $\tilde{\Rs} \le \Rs$
   rows are deleted from  $\deltav$, then the remaining matrix satisfies $\CR{\nfactrue}{\Rs-\tilde{\Rs}}$
   \item[(c)] Assume that $\CR{\nfactrue}{\Rs}$ holds for $\deltav$ and let $\deltav^\emptyset$ be the matrix after some or all zero rows are removed from $\deltav$.
   Then, $\CR{\nfactrue}{\Rs}$ holds for $\deltav^\emptyset$ as well.
	\end{itemize}
\end{lem}

\paragraph{Proof of Theorem~\ref{Crule357}.} %

Any matrix $\facloadtilde \in \ThetaD{\delta}$ has the same non-zero rows as $\deltav$.
Hence, if $\CR{\nfactrue}{\Rs}$ does not hold for $\deltav$, then it also does not hold for any $\facloadtilde \in \ThetaD{\delta}$.
According to Theorem 3.4.\ by~\citet{sat:stu} with their rotation $\rotG$ being the identity, this implies that $\RD{\nfactrue}{\Rs}$ is violated for all $\facloadtilde \in \ThetaD{\delta}$.
This proves part (a).

We prove part~(b) by induction. If $\CR{1}{\Rs}$ holds for a $\dimy \times 1$
sparsity vector $\deltav$, then
at least  $m_1 \geq 2  + \Rs$ elements of $\deltav$ are different from 0. It trivially
follows that all $\facload \in \ThetaD{\delta}$ have exactly the
same number of non-zero elements. After deleting $\Rs$ elements,
two subvectors with at least one non-zero elements can be formed and
$\RD{1}{\Rs}$ is satisfied for all
$\facloadtilde \in \ThetaD{\delta}$.
For any $\nfactrue \geq 2$, assume that part~(b) of Theorem~\ref{Crule357} holds
for $\nfactrue-1$ and
that the counting rule $\CR{\nfactrue}{\Rs}$ holds for an
$\dimy \times \nfactrue$ ordered GLT  sparsity
matrix $\deltav$.
A suitable  permutation of the rows of $\deltav$ yields:
\begin{eqnarray*}
	\Pim_r \deltav =  \left( \begin{array}{cc}
		\deltav^c  & \bfz \\
		\deltav^b  & \deltav^A \\
	\end{array} \right),
\end{eqnarray*}
where $\deltav^A$ is a GLT sparsity matrix with $\nfactrue-1$ columns, $\deltav^b$ and $\deltav^c$ are column vectors, and $\deltav^c$ contains $d_1 \geq1$ non-zero elements and no zero elements. %
According to Lemma~\ref{lemCRrules}(a) and (c), $\deltav^A$ satisfies $\CR{\nfactrue-1}{\Rs}$ and the first column $((\deltav^c)^\top, (\deltav^b)^\top)^\top$ satisfies $\CR{1}{\Rs}$.
Consequently, $\deltav^b$ contains at least $2+s-d_1$ non-zero elements.
Let  $\facloadtrue \in \ThetaD{\delta}$ be an ordered GLT matrix. %
If the same
$\Rs$ rows are deleted from  $\Pim_r \deltav$ and  $\Pim_r \facloadtrue$, we obtain the following matrices:
\begin{eqnarray*}
	\tilde \deltav =  \left( \begin{array}{cc}
		\tilde \deltav^c  & \bfz \\
		\tilde \deltav^b  & \tilde \deltav^A \\
	\end{array} \right), \qquad
	\tilde \facloadtrue=  \left( \begin{array}{cc}
		\tilde \facloadtrue^c  & \bfz \\
		\tilde \facloadtrue^b  & \tilde \facloadtrue^A \\
	\end{array} \right), \qquad
\end{eqnarray*}
where $0 \leq \Rs_1 \leq \min(d_1, \Rs)$ non-zero elements are deleted
from the vector $\deltav^c$  and $d_1-s_1$ non-zero elements remain in
the vectors  $\tilde \deltav^c$ and $\tilde \facloadtrue^c$,
while  the vectors $\tilde \deltav^b $ and  $\tilde \facloadtrue^b$ contain
$d_2 \geq \max (0, 2 +
\Rs_1-d_1)$ non-zero elements.
Since we removed $\Rs-\Rs_1$ rows from $\deltav^A$, according to Lemma~\ref{lemCRrules}(b), the sparsity matrix
$\tilde \deltav^A$ satisfies $\CR{\nfactrue-1}{\Rs_1}$ and hence, $\tilde \facloadtrue^A$ obeys
$\RD{\nfactrue-1}{\Rs_1}$ except for a set of measure 0.
We proceed with those matrices
$\tilde \facloadtrue^A$  where $\RD{\nfactrue-1}{\Rs_1}$ holds.
If further $\Rs_1$ rows are deleted from  $\tilde \facloadtrue^A$, then a matrix  results
which contains  two sub matrices $\Am_1$ and  $\Am_2$ of rank $\nfactrue-1$.
Let the $\Rs_1 \times (\nfactrue-1)$  matrix $\Bm$ contain the rows that were deleted from of $\tilde \facloadtrue^A$.
If the same rows are deleted from  $\tilde \facloadtrue^b$, then
the vector  $\bm$ containing the deleted elements has at least $\max( 0, 2 -(d_1 - \Rs_1))$
non-zero elements.
Next, we consider three cases. First, if $d_1 - \Rs_1 \geq 2$, then we use two of the
$d_1-s_1$ non-zero elements of  $\tilde \deltav^c$ to define following submatrices of
$\facloadtrue$:
\begin{eqnarray}
	\left( \begin{array}{cc}
		\tilde \deltav^c_{i_1} & \bfz \\
		\times & \Am_1 \\
	\end{array}
	\right), \quad      \left( \begin{array}{cc}
		\tilde \deltav^c_{i_2} & \bfz \\
		\times   & \Am_2 \\
	\end{array}
	\right).
\end{eqnarray}
Both matrices obviously have rank $\nfactrue$.
Second, if  $d_1 - \Rs_1 =1$, then we use the only
non-zero element of  $\tilde \deltav^c$ and one of the non-zero elements of $\bm$,
denoted by $ \bm_{i_2}$,
and the corresponding row $\Bm_{i_2,\bullet}$ of $\Bm$
to define
following submatrices of
$\facloadtrue$:
\begin{eqnarray}
	\left( \begin{array}{cc}
		\tilde \deltav^c_{i_1} & \bfz \\
		\times & \Am_1 \\
	\end{array}
	\right), \quad     \left( \begin{array}{cc}
		\bm_{i_2} & \Bm_{i_2,\bullet} \\
		\times   & \Am_2 \\
	\end{array}
	\right).
\end{eqnarray}
The first matrix obviously has rank $\nfactrue$.
The rank of the second matrix is  at least equal to $\rank{\Am_2}=\nfactrue-1$.
The row vector $\Bm_{i_2,\bullet}$ contains $0 \leq d_3\leq \nfactrue-1$ non-zero elements which
take
arbitrary values in $\mathbb{R}$. Hence the set of matrices where
the row $(\bm_{i_2} \,\, \Bm_{i_2,\bullet})$ is linearly dependent of the other rows and rank deficiency
occurs has measure zero.
Finally, if  $d_1 - \Rs_1 =0$, then we use two of the at least two non-zero elements in $\bm$,
denoted by $ \bm_{i_1}$ and $ \bm_{i_2}$,
and the corresponding rows $\Bm_{i_1,\bullet}$ and $\Bm_{i_2,\bullet}$ of $\Bm$ to define
following submatrices of
$\facloadtrue$:
\begin{eqnarray}
	\left( \begin{array}{cc}
		\bm_{i_1} & \Bm_{i_1,\bullet}\\
		\times & \Am_1 \\
	\end{array}
	\right), \quad      \left( \begin{array}{cc}
		\bm_{i_2} & \Bm_{i_2,\bullet} \\
		\times   & \Am_2 \\
	\end{array}
	\right).
\end{eqnarray}
Using the same argument as above, both matrices are
of rank $\nfactrue$ except for a set of measure 0.
This proves that $\RD{\nfactrue}{\Rs}$ holds for all GLT matrices $\facloadtrue \in \ThetaD{\delta}$
except for a set of measure 0.
The counting rule  $\CR{\nfactrue}{\Rs}$ is
invariant to signed permutations of $\deltav$.
Therefore, if $\CR{\nfactrue}{\Rs}$ implies $\RD{\nfactrue}{\Rs}$
for an ordered GLT matrix $\facloadtrue \in \ThetaD{\delta}$, then this holds for
all signed permutations  $\facloadtilde =  \facloadtrue \bP _{\pm} \bP _{\rho}$  of $\facloadtrue$.
This completes the proof of part~(b), since the set where $\RD{\nfactrue}{\Rs}$ does not hold is a finite
union of sets of measure 0.

\paragraph*{Proof of Corollary~\ref{Lemma1}.}
The three conditions in Corollary~\ref{Lemma1} follow immediately from Theorem~\ref{Crule357}.
The  $(j,l)$th coefficient  of  the matrix on the left hand of (\ref{checkNC2}) is given  by
$ d_j + \sum_{i=1}^\dimy \delta _{ij} (1-\delta _{il} )$,  where $d_j=\sum_{i=1}^{m} \delta _{ij}$ is the total
number of non-zero indicators in column $j$.
The diagonal elements ($j= l$) are equal to $d_j$ (since $\delta _{ij} (1-\delta _{ij} )=0$) and check
if  each column contains at least  $2+\Rs$ %
non-zero indicators.
The off-diagonal elements ($j \neq l$) count the number of nonzero rows in column $j$ and $l$.
Hence, the matrix on the right hand side of (\ref{checkNC2}) has diagonal elements equal to
$2+\Rs$ and off-diagonal elements equal to $4+\Rs$ .
The column vector $\deltav^\star$  in (\ref{checkNC1}) is equal to the
number of non-zero indicators in each row.
Hence, (\ref{checkNC1})  verifies if the total number of  nonzero rows of $\deltav$ is at
least equal to   $2\nfactrue + \Rs$. %
Finally, (\ref{checkNC3}) verifies if each submatrix of $\nfactrue-1$ columns has at least  $2\nfactrue -1$ nonzero rows. %
The $j$th column of the matrix $\deltav^\star$ appearing in (\ref{checkNC3})
is the number of non-zero indicators in each row
of the submatrix $\deltav_{-j}$ excluding the $j$th column. The matrix $ \indic{\deltav^\star >0}$
indicates nonzero rows  in $\deltav_{-j}$ and the $j$th  element  of  the row vector
$\ones_{1 \times \dimy} \cdot    \indic{\deltav^\star >0}$ counts the number of
nonzero rows  in $\deltav_{-j}$.

\paragraph*{Proof of Theorem~\ref{theoverGLT}.}
First, we prove further properties of the spurious factor matrix $\Mm_s$
in  representation (\ref{decover}) beyond the characterization given in\citet[Theorem~1]{tum-sat:ide}.
More specifically, we  show that the spurious cross-covariance
matrix  $\Mm_s \trans{\Mm_s} = \Dm_s$
is equal to a
diagonal matrix  of rank $s$, with  $s$ nonzero entries $d_{n_1}, \ldots, d_{n_s}$  in rows
$n_1, \ldots, n_\nspu$.   From $ \rank{\facload_k  \Tm_k}=\min(\rank{\facload_k} , \rank{\Tm_k})=
\nfactrue+s$, we obtain  that
$\Mm_s$  has full column rank $\rank{\Mm_s}=s$. Therefore
$ \rank{\Dm_s} =\rank{\Mm_s}=s$  and  only  $\nspu$ diagonal elements $d_{n_1},
\ldots, d_{n_s}$ of $\Dm_s$  in  rows $n_1, \ldots, n_\nspu$ are different from 0.
It is straightforward to show that the  matrix $\Mm_s $ has exactly the same $s$ nonzero
rows $n_1, \ldots, n_\nspu$ as $\Dm_s$: using  %
for each row $\Mm_{i,\cdot}$ of  $\Mm_s $ that  $\Mm_{i,\cdot} \trans{\Mm_{i,\cdot}}=\|\Mm_{i,\cdot}\|_2^2=d_i$, %
it follows for any  $i \neq \{n_1, \ldots, n_\nspu\}$  that $\|\Mm_{i,\cdot}\|_2^2=0$
and, therefore, $\Mm_{i,\cdot}=\bfz$, whereas the remaining rows with $i \in \{n_1, \ldots, n_\nspu\}$ are nonzero since $\|\Mm_{i,\cdot}\|_2^2 >0$.
The  submatrix $\Mm _0$ of nonzero rows in $\Mm_s $ satisfies $ \Mm _0 \trans{\Mm _0}= \Dm_0^2$ with $\Dm_0 ^2=\Diag{d_{n_1}, \ldots, d_{n_s} }$ being a diagonal matrix of rank $s$.
It follows that  $\Dm_0^{-1} \Mm_s  \trans{\Dm_0^{-1}  \Mm_s}   = \identm$,
hence  $ \Dm_0^{-1} \Mm_s  = \Qm $
for any arbitrary rotation matrix $\Qm $ of rank $s$. Therefore:
\begin{eqnarray} \label{decmmm}
	\Mm _0 =  \Dm_0 \Qm , \qquad  \Dm_0= \Diag{d_{n_1}, \ldots, d_{n_s} }^{1/2}.
\end{eqnarray}
Let $\facload_k ^\star$, $\Vare_k^\star$, $\Mm_s ^\star$,
$\facloadtrue ^\star$, and $\Varetrue^\star $,
be the matrices that result from deleting  the  rows $n_1, \ldots, n_s$
(and for $\Vare_k$ and $\Varetrue$ also the corresponding columns)
from  the matrices $\facload_k$,  $\Vare_k$, $\Mm_s $, $\facloadtrue$,  and $\Varetrue$
in representation (\ref{decover}).
Condition $\RD{\nfactrue}{1+\tumS}$ for $\facloadtrue$
implies that $\facloadtrue ^\star$ satisfies condition $\RD{\nfactrue}{1}$
and the  variance decomposition
	$\Vary ^\star=    \facloadtrue ^\star \trans{(\facloadtrue ^\star) } + \Varetrue^\star $
is unique.   Since $\Mm_s ^\star =\bfzmat$,   we obtain from (\ref{decover})  that
\begin{eqnarray} \label{eqeq}
	\facload_k^\star  \Tm_k =  \left(\begin{array}{cc}
		\facloadtrue ^\star & \bfzmat
	\end{array} \right) ,    \qquad   \Vare_k ^\star =   \Varetrue ^\star,
\end{eqnarray}
hence $\facload_k^\star \trans{(\facload_k^\star)} = \facloadtrue ^\star \trans{(\facloadtrue ^\star)}$
and  $\facload_k^\star$ has reduced rank   $\rank{\facload_k^\star }=
\rank{\facloadtrue ^\star} = \nfactrue$.

So far, the results are valid for any rotational strategy.
We turn now to the case that
$\facloadtrue$ is a GLT matrix and $\facload_k $ in the EFA model is constrained
to be a unordered GLT matrix of rank $\nfactrue+s$,  with non-zero factor loadings
in $\nfactrue+s$ different pivot rows in the set
$\{\tilde{l}_1, \ldots, \tilde{l}_{\nfactrue+s} \}$.
A necessary condition for  $\facload_k^\star$ to have reduced rank $\nfactrue$
is that the  deleted rows $n_1, \ldots, n_s$
are equal to  $s$  of these pivot rows.
The remaining rows in the corresponding $s$  columns of $\facload_k^\star$ are linearly
independent, except for a set of measure zero,
and $\facload_k^\star$ has rank $\nfactrue$, iff these
columns  are zero.

A signed permutation ${\Pm}=\bP _{\pm} \bP _{\rho} $ is be used to reorder the
columns of $\facload_k $ on the left hand side of (\ref{decover}). The first
$\nfactrue$ columns of $\tilde{\facload}_k=\facload_k {\Pm} $  correspond
to the nonzero columns  of $\facload_k^\star$
and are arranged
such that the resulting $\dimy \times \nfactrue$  matrix $\facload_r$ (and the
corresponding matrix $\facload^\star_r$) is an ordered GLT matrix with
the   pivots
$\tilde{l}_1 < \ldots < \tilde{l}_{\nfactrue}$ that remain after deleting the $s$ rows.
The last $s$ columns of $\tilde{\facload}_k$ are
arranged such that they are an ordered GLT matrix  with  pivots $n_1 < \ldots < n_s$.
Changing  the rotation  $\Tm_k$ in (\ref{decover}) accordingly,
we obtain on the left hand side $\facload_k \Tm_k= \tilde{\facload}_k \Tm^\star$, where
$\Tm^\star = \Pm \Tm_k$. Split $\Tm^\star$ in the following way:
\begin{eqnarray*}
	\Tm^\star = \left(
	\begin{array}{cc}
		\Tm^\star_1 & \Tm^\star_{3}  \\
		\trans{(\Tm^\star_{3})}  &  \Tm^\star_2 \\
	\end{array}
	\right).
\end{eqnarray*}
From (\ref{eqeq}) we obtain:
\begin{eqnarray*}
	\facload^\star_k \Tm_k= \tilde{\facload}^\star_k \Tm^\star=
	\left(\begin{array}{cc}
		\facload^\star_r \Tm^\star_1  & \facload^\star_r \Tm^\star_3
	\end{array} \right)=
	\left(\begin{array}{cc}
		\facloadtrue ^\star & \bfzmat
	\end{array} \right).
\end{eqnarray*}
Since $\facload^\star_r $ has full column rank, we obtain
from $\facload^\star_r \Tm^\star_3 = \bfzmat$
by left multiplication with $(\trans{(\facload^\star_r)} \facload^\star_r) ^{-1}
\trans{(\facload^\star_r)}$
that $\Tm^\star_3= \bfzmat$.
Furthermore, $\facload^\star_r \Tm^\star_1 = \facloadtrue ^\star$.
Application of Theorem~\ref{theGLT} to the GLT matrix $\facload^\star_r  $,
which satisfies $ \facload^\star_r \trans{( \facload^\star_r)}
=\facload_k^\star \trans{(\facload_k^\star)} =   \facloadtrue ^\star \trans{(\facloadtrue ^\star)}$,
yields $ \facload_r ^\star  =  \facloadtrue ^\star  $.
Therefore,  the pivots
$\tilde{l}_1 , \ldots , \tilde{l}_{\nfactrue}$ are equal to the pivots
${l}_1 , \ldots , {l}_{\nfactrue}$ of $\facloadtrue$.
Furthermore, we obtain $\Tm^\star_1 = \identy{\nfactrue}$.

Consider now the $s$ rows $\facload_{k}^d$ that were deleted from $\facload_k$ and
denote the corresponding rows
in the reordered
matrix $\tilde{\facload}_k$ as $\tilde{\facload}_k^d$.
The
first $\nfactrue$ columns of $\tilde{\facload}_k^d$ are the rows $\facload_{r}^d$
deleted from $\facload_r$
to define $\facload^\star_r$ and the
last $s$  columns  are a lower triangular matrix $\Lm$ with  pivot elements
$n_1, \ldots , n_s$  on the main diagonal.
From $ \facload_{k}^d \Tm_k= \tilde{\facload}_k^d \Tm^\star $, we obtain
\begin{eqnarray*}
	\left(\begin{array}{cc}
		\facload_{r}^d  & \Lm \Tm^\star_2
	\end{array} \right)=
	\left(\begin{array}{cc}
		\facloadtrue^d  & \Mm _0
	\end{array} \right).
\end{eqnarray*}
Hence, the deleted rows $\facload_{r}^d$ are identical to the rows $\facloadtrue^d$
deleted from $\facloadtrue$ and
the first $\nfactrue$ columns of $\tilde{\facload}_k$  are
equal to $\facloadtrue $.
Since $  \Lm \Tm^\star_2 \trans{(\Tm^\star_2)}  \trans{\Lm} =
\Lm \trans{\Lm}  =
\Mm _0  \trans{\Mm _0} = \Dm_0^2 $,
where $\Dm_0$ is a diagonal matrix, it follows that $\Lm $ is
a diagonal matrix and equal to $\Dm_0$.
Hence, in the GLT framework, the last $s$ columns of $\tilde{\facload}_k$,
corresponding the spurious factor matrix,
can be represented as a spurious ordered GLT matrix $\Mm^{\Lambda}_s$, where the pivot rows
are equal to $\Lm $.
This completes the proof:
\begin{eqnarray*}
	\left(\begin{array}{cc}
		\facloadtrue  & \Mm ^{\Lambda}_s
	\end{array} \right)  =
	\tilde{\facload}_k  = \facload_k {\Pm} =  \facload_k \bP _{\pm} \bP _{\rho} .
\end{eqnarray*}

\end{document}